\documentclass[envcountsame]{llncs}

\usepackage{times}

\usepackage{times}
\usepackage{amsmath}
\usepackage{amsfonts}
\usepackage{amssymb}
\usepackage{mathpartir} 
\usepackage{xcolor}
\usepackage{xspace} 
\usepackage[pdftex]{graphicx}
\usepackage{fancybox}
\usepackage{url}
\usepackage{multirow}
\usepackage{colortbl}
\usepackage{framed}
\usepackage{mathtools}
\usepackage{enumitem}
\usepackage{wrapfig}
\usepackage{hyperref}
\usepackage{version}

\includeversion{rep}
\excludeversion{conf}

\newcommand{\dblind}[1]{\colorbox{black!85}{\textcolor{white}{{\texttt{redacted}}}}}

\newcommand{\mytitle}{Solving Quantified Bit-Vectors \\ using Invertibility Conditions}

\hypersetup{
  pdfborder={0 0 0},
  colorlinks=true,
  linkcolor=blue,
  urlcolor=blue,
  citecolor=blue,
  pdftitle=\mytitle
}

\excludeversion{longv}
\includeversion{shortv}

\newcommand\xqed[1]{%
  \leavevmode\unskip\penalty9999 \hbox{}\nobreak\hfill
  \quad\hbox{#1}}
\newcommand\exqed{\xqed{$\triangle$}}

\newcommand{\cvc}{CVC4\xspace}
\newcommand{\ziii}{Z3\xspace}
\newcommand{\boolector}{Boolector\xspace}
\newcommand{\qiiib}{Q3B\xspace}
\newcommand{\yices}{Yices\xspace}

\newcommand{\rem}[1]{\textcolor{red}{[#1]}}

\renewcommand{\vec}[1]{\ensuremath{\boldsymbol{#1}}\xspace}

\newcommand{\sig}{\ensuremath{\Sigma}\xspace}
\newcommand{\sigs}{\ensuremath{\sig^s}\xspace}
\newcommand{\sigf}{\ensuremath{\sig^f}\xspace}
\newcommand{\sigbv}{\ensuremath{\sig_{BV}}\xspace}

\newcommand{\tbv}{\ensuremath{T_{BV}}\xspace}

\newcommand{\teq}{\ensuremath{\approx}\xspace}
\newcommand{\tneq}{\ensuremath{\not\teq}\xspace}

\newcommand{\sort}{\ensuremath{\sigma}\xspace}
\newcommand{\sorts}{\ensuremath{S}\xspace}
\newcommand{\sorti}{\ensuremath{\sort^\I}\xspace}
\newcommand{\sortbv}[1]{\ensuremath{\sort_{[#1]}}\xspace}
\newcommand{\bv}[2]{\ensuremath{#1_{[#2]}}\xspace}

\newcommand{\vars}{\ensuremath{X}\xspace}
\newcommand{\varss}[1]{\ensuremath{\vars_{#1}}\xspace}
\newcommand{\varsbv}[1]{\ensuremath{\vars_{[#1]}}\xspace}
\newcommand{\choicef}{\ensuremath{\varepsilon}\xspace}
\newcommand{\choice}[2]{\ensuremath{\choicef #1.\,#2}\xspace}
\newcommand{\binda}[2]{\ensuremath{\forall #1.\,#2}\xspace}
\newcommand{\binde}[2]{\ensuremath{\exists #1.\,#2}\xspace}
\newcommand{\fv}[1]{\ensuremath{{FV}\kern-.15em(#1)}\xspace}

\newcommand{\I}{\ensuremath{{\mathcal{I}}}\xspace}
\newcommand{\J}{\ensuremath{{\mathcal{J}}}\xspace}
\newcommand{\Is}{\ensuremath{{I}}\xspace}

\newcommand{\lits}[1]{\ensuremath{\mathrm{Lit}(#1)}\xspace}

\newcommand{\bool}{\ensuremath{\mathsf{Bool}}\xspace}

\newcommand{\cegqi}[1]{\ensuremath{\mathsf{CEGQI}_{#1}}}

\newcommand{\selfun}{\ensuremath{\mathcal{S}}}

\newcommand{\rel}{\ensuremath{\bowtie}\xspace}

\newcommand{\selfunbv}[1]{\ensuremath{\mathcal{S}^{BV}_{#1}}\xspace}
\newcommand{\project}[1]{\ensuremath{\mathsf{project}_{#1}}\xspace}
\newcommand{\solvelitf}{\ensuremath{\mathsf{solve}}\xspace}
\newcommand{\solvelit}[1]{\ensuremath{\solvelitf(#1)}\xspace}
\newcommand{\linearize}[3]{\ensuremath{\linearizef(#1, #2, #3)}\xspace}
\newcommand{\linearizef}{\ensuremath{\mathsf{linearize}}}
\newcommand{\choosef}{\ensuremath{\mathsf{choose}}}
\newcommand{\getinversef}{\ensuremath{\mathsf{getInverse}}\xspace}
\newcommand{\getinverse}[2]{\ensuremath{\getinversef(#1, #2)}\xspace}
\newcommand{\getscf}{\ensuremath{\mathsf{getIC}}\xspace}
\newcommand{\getsc}[2]{\ensuremath{\getscf(#1, #2)}\xspace}

\newcommand{\op}{\ensuremath{\diamond}\xspace}

\newcommand{\bvaddf}{\ensuremath{+}\xspace}
\newcommand{\bvandf}{\ensuremath{\mathrel{\&}}\xspace}
\newcommand{\bvashrf}{\ensuremath{\mathop{>\kern-.3em>_a}}\xspace}
\newcommand{\bvconcatf}{\ensuremath{\circ}\xspace}
\newcommand{\bvlshrf}{\ensuremath{\mathop{>\kern-.3em>}}\xspace}
\newcommand{\bvmulf}{\ensuremath{\cdot}\xspace}
\newcommand{\bvorf}{\ensuremath{\mid}\xspace}
\newcommand{\bvsgef}{\ensuremath{\ge_s}\xspace}
\newcommand{\bvsgtf}{\ensuremath{>_s}\xspace}
\newcommand{\bvshlf}{\ensuremath{\mathop{<\kern-.3em<}}\xspace}
\newcommand{\bvslef}{\ensuremath{\le_s}\xspace}
\newcommand{\bvsltf}{\ensuremath{<_s}\xspace}
\newcommand{\bvudivf}{\ensuremath{\div}\xspace}
\newcommand{\bvugef}{\ensuremath{\geq_u}\xspace}
\newcommand{\bvugtf}{\ensuremath{>_u}\xspace}
\newcommand{\bvultf}{\ensuremath{<_u}\xspace}
\newcommand{\bvulef}{\ensuremath{\leq_u}\xspace}
\newcommand{\bvuremf}{\ensuremath{\bmod}\xspace}
\newcommand{\bvnegf}{\ensuremath{-}\xspace}
\newcommand{\bvnotf}{\ensuremath{{\sim}\,}\xspace}

\newcommand{\booland}[2]{\ensuremath{#1\,\wedge\,#2}\xspace}
\newcommand{\boolnot}[1]{\ensuremath{\neg #1}\xspace}
\newcommand{\boolor}[2]{\ensuremath{#1\,\vee\,#2}\xspace}
\newcommand{\bvadd}[2]{\ensuremath{#1 \bvaddf #2}\xspace}
\newcommand{\bvand}[2]{\ensuremath{#1 \bvandf #2}\xspace}
\newcommand{\bvashr}[2]{\ensuremath{#1 \bvashrf #2}\xspace}
\newcommand{\bvconcat}[2]{\ensuremath{#1 \bvconcatf #2}\xspace}
\newcommand{\bvextract}[3]{\ensuremath{#1[#2:#3]}\xspace}
\newcommand{\bvlshr}[2]{\ensuremath{#1 \bvlshrf #2}\xspace}
\newcommand{\bvmul}[2]{\ensuremath{#1 \bvmulf #2}\xspace}
\newcommand{\bvneg}[1]{\ensuremath{\bvnegf#1}\xspace}
\newcommand{\bvnot}[1]{\ensuremath{\bvnotf\!#1}\xspace}
\newcommand{\bvor}[2]{\ensuremath{#1 \bvorf #2}\xspace}
\newcommand{\bvsge}[2]{\ensuremath{#1 \bvsgef #2}\xspace}
\newcommand{\bvsgt}[2]{\ensuremath{#1 \bvsgtf #2}\xspace}
\newcommand{\bvshl}[2]{\ensuremath{#1 \bvshlf #2}\xspace}
\newcommand{\bvsle}[2]{\ensuremath{#1 \bvslef #2}\xspace}
\newcommand{\bvslt}[2]{\ensuremath{#1 \bvsltf #2}\xspace}
\newcommand{\bvsub}[2]{\ensuremath{#1 - #2}\xspace}
\newcommand{\bvudiv}[2]{\ensuremath{#1 \bvudivf #2}\xspace}
\newcommand{\bvuge}[2]{\ensuremath{#1 \bvugef #2}\xspace}
\newcommand{\bvugt}[2]{\ensuremath{#1 \bvugtf #2}\xspace}
\newcommand{\bvule}[2]{\ensuremath{#1 \bvulef #2}\xspace}
\newcommand{\bvult}[2]{\ensuremath{#1 \bvultf #2}\xspace}
\newcommand{\bvurem}[2]{\ensuremath{#1 \bvuremf #2}\xspace}
\newcommand{\dist}[2]{\ensuremath{#1 \tneq #2}\xspace}
\newcommand{\equal}[2]{\ensuremath{#1 \teq #2}\xspace}
\newcommand{\imp}[2]{\ensuremath{#1\hspace{0.1em}\Rightarrow\hspace{0.1em}#2}\xspace}

\newcommand{\true}{\ensuremath{\top}\xspace}

\newcommand{\maxs}{\ensuremath{\text{max}_s}\xspace}
\newcommand{\mins}{\ensuremath{\text{min}_s}\xspace}

\newcommand{\bwf}{\ensuremath{\kappa}\xspace}
\newcommand{\bw}[1]{\ensuremath{\bwf(#1)}\xspace}

\newcommand{\cfgeqb}{\ensuremath{\mathbf{b}}}
\newcommand{\cfgeqs}{\ensuremath{\mathbf{s}}}
\newcommand{\cfgkeep}{\ensuremath{\mathbf{k}}}
\newcommand{\cfgmval}{\ensuremath{\mathbf{m}}}

\newcommand{\solver}{\ensuremath{\bf{cegqi}}\xspace}
\newcommand{\solvercfg}[1]{\ensuremath{\bf{\solver}_{#1}}\xspace}

\newcommand{\dpllt}{DPLL$(T)$\xspace}

\newcommand{\ic}{\phi_\mathrm{c}}
\newcommand{\limpl}{\Rightarrow}
\newcommand{\lequiv}{\Leftrightarrow}

\newcolumntype{L}{>{$}l<{$}} 
\newcolumntype{R}{>{$}r<{$}} 
\newcolumntype{C}{>{$}c<{$}} 

\begin{document}

\title{\mytitle}

\titlerunning{\mytitle}

\author{%
  Aina Niemetz\inst{1}
  \and
  Mathias Preiner\inst{1}
  \and
  Andrew Reynolds\inst{2}
  \and\\
  Clark Barrett\inst{1}
  \and
  Cesare Tinelli\inst{2}
}

\institute{
{Stanford University}
\and
{The University of Iowa}
}

\maketitle

\begin{abstract}
  We present a novel approach for solving quantified bit-vector formulas
  in Satisfiability Modulo Theories (SMT)
  based on computing symbolic inverses of bit-vector operators.
  We derive conditions that precisely characterize when bit-vector constraints
  are invertible for a representative set of bit-vector operators
  commonly supported by SMT solvers.
  We utilize syntax-guided synthesis techniques
  to aid in establishing these conditions
  and verify them independently by using several SMT solvers.
  We show that
  invertibility conditions can be embedded into quantifier instantiations
  using Hilbert choice expressions, and give experimental evidence that a
  counterexample-guided approach for quantifier instantiation
  utilizing these techniques leads to performance improvements with respect
  to state-of-the-art solvers for quantified bit-vector constraints.
\end{abstract}

\pagestyle{plain}

\section{Introduction}

  Many applications in hardware and software verification
  rely on Satisfiability Modulo Theories (SMT) solvers
  for bit-precise reasoning.
  In recent years,
  the quantifier-free fragment of the theory of fixed-size bit-vectors
  has received a lot of interest, as witnessed by the number of
  applications that generate problems in that fragment and  
  by the high, and increasing, number of solvers that
  participate in the corresponding division of the annual SMT competition.
  Modeling properties of programs and circuits,
  e.g., universal safety properties and program invariants,
  however,
  often requires the use of \emph{quantified} bit-vector formulas.
  Despite a multitude of applications,
  reasoning efficiently about such formulas is still 
  a challenge in the automated reasoning community.

  The majority of solvers that support quantified bit-vector logics
  employ instantiat\-ion-based
  techniques~\cite{WintersteigerHM13,ReynoldsDKTB15,yicesef,PreinerNB17},
  which aim to find conflicting ground instances of quantified formulas.
  For that, it is crucial to select good instantiations for the universal variables,
  or else the solver may be overwhelmed by the number of ground instances
  generated.
  For example,
  consider a quantified formula
  $\psi = \binda{x}{(\dist{\bvadd{x}{{s}}}{t})}$
  where $x$, $s$ and $t$ 
  denote bit-vectors of size 32.
  To prove that $\psi$ is unsatisfiable
  we can instantiate $x$ with all $2^{32}$ possible bit-vector values.
  However, ideally, we would like to find a proof
  that requires much fewer instantiations.
  In this example, if we instantiate $x$ with the symbolic term $t - s$
  (the inverse of $\equal{\bvadd{x}{s}}{t}$ when solved for $x$),
  we can
  immediately conclude that $\psi$ is unsatisfiable since
  $\bvadd{(\bvsub{t}{s})}{{s} \tneq t}$ simplifies to false.

  Operators in the theory of bit-vectors are not
  always invertible. However, we observe
  it is possible to identify quantifier-free conditions that 
  precisely \emph{characterize} when they are.
  We do that for a representative set of operators in the standard theory 
  of bit-vectors supported by SMT solvers.
  For example, we have proven that 
  the constraint
  \equal{\bvmul{x}{s}}{t} is solvable for $x$ if and only if 
  \equal{\bvand{(\bvor{\bvneg{s}}{s})}{t}}{t} is satisfiable.
  %
  Using this observation, 
  we develop a novel approach for solving quantified
  bit-vector formulas that utilizes invertibility conditions
  to generate symbolic instantiations.
  We show that invertibility conditions can be embedded
  into quantifier instantiations using Hilbert choice functions
  in a sound manner.
  This approach has compelling advantages
  with respect to previous approaches, which we demonstrate
  in our experiments.  
  
  More specifically, this paper makes the following \emph{contributions}.  

  \begin{itemize}
  \item We derive and present invertibility conditions
  for a representative set of bit-vector operators
  that allow us to model all bit-vector constraints
  in SMT-LIB~\cite{SMTLib2010}.
  \item
  We provide details on how invertibility conditions can be automatically
  synthesized using 
  syntax-guided synthesis (SyGuS)~\cite{AlurBJMRSSSTU13} techniques,
  and make public 162 available 
  challenge problems for SyGuS solvers that are encodings of this task.
  \item
  We prove that our approach 
  can efficiently reduce a class of quantified formulas, 
  which we call \emph{unit linear invertible}, 
  to quantifier-free constraints.
  \item
  Leveraging invertibility conditions,
  we implement a novel quantifier instantiation scheme 
  as an extension of the 
  SMT solver \cvc~\cite{CVC4},
  which shows improvements with respect to state-of-the-art 
  solvers for quantified bit-vector constraints.
  \end{itemize}

  \noindent\emph{Related work}
  Quantified bit-vector logics
  are currently supported by the SMT solvers
  \boolector~\cite{boolector2014},
  \cvc~\cite{CVC4},
  \yices~\cite{yices2014},
  and \ziii~\cite{Z3}
  and a Binary Decision Diagram (BDD)-based tool called \qiiib~\cite{q3b2016}.
  Out of these, only \cvc and \ziii provide support
  for combining quantified bit-vectors with other theories,
  e.g., the theories of arrays or real arithmetic.
  Arbitrarily nested quantifiers are handled by all but \yices,
  which only supports bit-vector formulas of the form
  $\binde{\vec{x}\forall \vec{y}}{Q[\vec{x}, \vec{y}]}$~\cite{yicesef}.
  For quantified bit-vectors, \cvc employs
  counterexample-guided quantifier instantiation (CEGQI)~\cite{ReynoldsDKTB15},
  where concrete models of a set of ground instances and
  the negation of the input formula (the counterexamples)
  serve as instantiations for the universal variables.
  In \ziii,
  model-based quantifier instantiation (MBQI)~\cite{GeM09}
  is combined
  with a template-based model finding procedure~\cite{WintersteigerHM13}.
  In contrast to \cvc,~\ziii not only relies on concrete counterexamples
  as candidates for quantifier instantiation
  but generalizes these counterexamples
  to generate symbolic instantiations
  by selecting ground terms with the same model value.
  \boolector employs a syntax-guided synthesis approach to synthesize
  interpretations for Skolem functions based on a set of ground instances of
  the formula, and uses a counterexample refinement loop
  similar to MBQI~\cite{PreinerNB17}.
  Other counterexample-guided approaches for quantified formulas in SMT solvers
  have been considered by Bj{\o}rner and Janota~\cite{bjornerplaying} and 
  by Reynolds et al.~\cite{DBLP:journals/fmsd/ReynoldsKK17},
  but they have mostly targeted quantified linear arithmetic and do not
  specifically address bit-vectors.
  Quantifier elimination for a fragment of bit-vectors
  that covers modular linear arithmetic has been recently addressed
  by John and Chakraborty~\cite{DBLP:journals/fmsd/JohnC16}, although
  we do not explore that direction in this paper.

\section{Preliminaries}

  %
  We assume the usual notions and terminology of many-sorted first-order logic
  with equality (denoted by $\teq$).
  Let \sorts be a set of \emph{sort symbols},
  and for every sort ${\sort \in \sorts}$
  let~\varss{\sort} be an infinite set of \emph{variables of sort \sort}.
  We assume that sets \varss{\sort} are pairwise disjoint
  and define \vars as the union of sets \varss{\sort}.
  Let \sig be a \emph{signature}
  consisting of
  a set
  $\sigs\!\subseteq \sorts$
  of sort symbols
  and
  a set
  $\sigf$
  of interpreted (and sorted) function symbols
  $f^{\sort_1 \cdots \sort_n \sort}$
  with arity $n \ge 0$ and $\sort_1, ..., \sort_n, \sort \in \sigs$.
  We assume that a signature \sig includes a Boolean sort \bool
  and the Boolean constants $\top$ (true) and $\bot$ (false).
  Let \I be a \emph{\sig-interpretation} that maps:
  each $\sort \in \sigs$ to a non-empty set \sorti
  (the \emph{domain} of \I),
  with $\bool^\I = \{ \top, \bot \}$;
  each $x \in \varss{\sort}$ to an element ${x^\I \in \sorti}$;
  and each $f^{\sort_1 \cdots \sort_n \sort} \in \sigf$ to a total function
  $f^\I\!\!: \sort_1^\I \times ... \times \sort_n^\I \to \sorti$ if $n > 0$,
  and to an element in \sorti if $n = 0$.
  If $x \in X_\sigma$ and $v \in \sigma^\I$, we denote by $\I[x \mapsto v]$
  the interpretation that maps $x$ to $v$ and is otherwise identical to $\I$.
  We use the usual inductive definition of a satisfiability relation $\models$
  between \sig-interpretations and \sig-formulas.
  
  We assume the usual definition of
  well-sorted terms, literals, and formulas as \bool terms with
  variables in \vars and symbols in \sig,
  and refer to them as \sig-terms, \sig-atoms, and so on.
  %
  %
  A \emph{ground} term/formula is a \sig-term/formula without variables.
  We define $\vec{x} = (x_1, ..., x_n)$ as a tuple of variables
  and write $Q\vec{x}\varphi$ with $Q \in \{ \forall, \exists \}$
  for a \emph{quantified} formula 
  $Qx_1 \cdots Qx_n \varphi$.
  We use \lits{\varphi} to denote the set of \sig-literals of 
  \sig-formula $\varphi$.
  For a \sig-term or \sig-formula $e$,
  we denote the \emph{free variables} of $e$ (defined as usual) as \fv{e} and
  use $e[\vec{x}]$ to denote that the variables in \vec{x} occur free in $e$.
  For a tuple of \sig-terms $\vec{t} = (t_1, ..., t_n)$,
  we write $e[\vec{t}]$ for the term or formula obtained from $e$
  by simultaneously replacing each occurrence of $x_i$ in $e$ by $t_i$.
  %
%
  %
  Given a \sig-formula $\varphi[x]$ with ${x \in \varss{\sort}}$,
  we use
  Hilbert's \emph{choice}
  operator
  \choicef~\cite{hilbert1934grundlagen}
  to describe \emph{properties} of $x$.
  We define a
  \emph{choice function}
  \choice{x}{\varphi[x]}
  as a term
  where $x$ is bound by \choicef.
  In every interpretation $\I$, 
  \choice{x}{\varphi[x]} denotes some value $v \in \sorti$ 
  such that $\I[x \mapsto v]$ satisfies $\varphi[x]$
  if such values exist,
  and denotes an arbitrary element of $\sorti$ otherwise.  
  This means that the formula
  $\binde{x}{\varphi[x]} \lequiv \varphi[\choice{x}{\varphi[x]}]$
  is satisfied by every interpretation.

  A \emph{theory}~$T$ is a pair $(\sig, \Is)$,
  where \sig is a signature
  and \Is is a non-empty class of \sig-interpretations
  (the \emph{models} of $T$)
  that is closed under variable reassignment,
  i.e.,
  every \sig-interpretation that only differs from an ${\I \in \Is}$
  in how it interprets variables is also in~\Is.
  A \sig-formula $\varphi$ is \emph{$T$-satisfiable} (resp.~\emph{$T$-unsatisfiable})
  if it is satisfied by some (resp.~no) interpretation in \Is;
  it is \emph{$T$-valid} if it is satisfied by all interpretations in \Is.
  \begin{rep}
  We say $T$ is a \emph{complete} theory if for all closed $\sig$-formulas $\varphi$,
  $\varphi$ is either $T$-valid or $T$-unsatisfiable.
  \end{rep}
  %
  A choice function \choice{x}{\varphi[x]} is \emph{($T$-)valid}
  if \binde{x}{\varphi[x]} is ($T$-)valid.
  We refer to a term~$t$ as \emph{\choicef-($T$-)valid}
  if all occurrences of choice functions in $t$ are ($T$-)valid.
  We will sometimes omit $T$ when the theory is understood from context.

  %
  We will focus on the theory $\tbv = (\sigbv, \Is_{BV})$
  of fixed-size bit-vectors as defined by the SMT-LIB 2 standard~\cite{SMTLib2010}.
  The signature $\sigbv$ includes a unique sort for each positive 
  bit-vector width $n$,
  denoted here as $\sortbv{n}$.
  Similarly,~\varsbv{n}
  is the set of \emph{bit-vector variables} of sort $\sortbv{n}$,
  and $\varss{BV}$ 
  is the union of all sets \varsbv{n}.
  We assume that \sigbv includes all \emph{bit-vector constants}
  of sort $\sortbv{n}$ for each $n$, represented as bit-strings.
  However, to simplify the notation we will sometimes denote them by the corresponding
  natural number in $\{0, \ldots, 2^{n-1}\}$.
  All interpretations $\I \in \Is_{BV}$ are identical except for the value
  they assign to variables. They interpret sort and function symbols as
  specified in SMT-LIB 2.
  All function symbols in $\sigf_{BV}$ are overloaded for every
  $\sortbv{n}\! \in \sigs_{BV}$.
  We denote
  a \sigbv-term (or \emph{bit-vector term}) $t$ of width $n$ as \bv{t}{n}
  when we want to specify its bit-width explicitly.
  %
  %
  We use $\text{max}_{s{[n]}}$ or $\text{min}_{s{[n]}}$
  for the \emph{maximum} or \emph{minimum signed value} of width $n$,
  e.g., $\text{max}_{s{[4]}} = 0111$ and $\text{min}_{s{[4]}}=1000$.
  The width of a bit-vector sort or term
  is given by the function \bwf,
  e.g., $\bw{\sortbv{n}} = n$ and $\bw{\bv{t}{n}} = n$.
  %

  Without loss of generality,
  we consider a restricted set of bit-vector function symbols
  (or \emph{bit-vector operators})
  $\sigf_{BV}$ as listed in Table~\ref{tab:bvops}.
  The selection of operators in this set is arbitrary
  but complete in the sense that
  it suffices to express
  all bit-vector operators defined in SMT-LIB 2.

\begin{table}[t]
  \centering
{%
  \renewcommand{\arraystretch}{1.2}%
  \begin{tabular}{l@{\hspace{1.6em}}l@{\hspace{1.6em}}l}
    \hline
    \textbf{Symbol} & \textbf{SMT-LIB Syntax} & \textbf{Sort} \\
    \hline

    \teq, \bvultf, \bvugtf, \bvsltf, \bvsgtf
      & =, bvult, bvugt, bvslt, bvsgt
      & $\sort_{[n]} \times \sort_{[n]} \to \bool$
      \\
    \bvnotf, \bvnegf
      & bvnot, bvneg
      & $\sort_{[n]} \to \sort_{[n]}$
      \\
    \bvandf, \bvorf, \bvshlf, \bvlshrf, \bvashrf
      & bvand, bvor, bvshl, bvlshr, bvashr
      & $\sort_{[n]} \times \sort_{[n]} \to \sort_{[n]}$
      \\
    \bvaddf, \bvmulf, \bvuremf, \bvudivf
      & bvadd, bvmul, bvurem, bvudiv
      & $\sort_{[n]} \times \sort_{[n]} \to \sort_{[n]}$
      \\
    \bvconcatf
      & concat
      & $\sort_{[n]} \times \sort_{[m]} \to \sort_{[n+m]}$
      \\
    \bvextract{}{u}{l}
      & extract
      & $\sort_{[n]} \to \sort_{[u-l+1]}$, $0 \le l \le u < n$
      \\
    \hline
  \end{tabular}%
}

  \vspace{1ex}
  \caption{Set of considered bit-vector operators with
  corresponding SMT-LIB 2 syntax.}
  \label{tab:bvops}
\end{table}

\section{Invertibility Conditions for Bit-Vector Constraints}
\label{sec:bvinversion}

  This section formally 
  introduces the concept of an invertibility condition
  and shows that such conditions can be used to construct
  symbolic solutions for a class of quantifier-free bit-vector constraints
  that have a linear shape.

  Consider a bit-vector literal \equal{\bvadd{x}{s}}{t}
  and assume that we want to solve for $x$.
  If the literal is \emph{linear} in $x$, that is, has only one
  occurrence of $x$, a general solution for $x$ is given by 
  the inverse of bit-vector addition over equality: $x = \bvsub{t}{s}$.
  Computing the inverse of a bit-vector operation,
  however,
  is not always possible.
  For example,
  for \equal{\bvmul{x}{s}}{t},
  an inverse always exists only if $s$ always evaluates to an odd bit-vector.
  Otherwise,
  there are values for $s$ and $t$
  where no such inverse exists, e.g., \equal{\bvmul{x}{2}}{3}.
  However, even if there is no unconditional inverse for the general case,
  we can identify the condition
  under which a bit-vector operation is invertible.
  For the bit-vector multiplication constraint \equal{\bvmul{x}{s}}{t}
  with $x \notin \fv{s} \cup \fv{t}$,
  the \emph{invertibility condition} for $x$ can be expressed by the formula
  \equal{\bvand{(\bvor{\bvneg{s}}{s})}{t}}{t}.

  \begin{definition}(Invertibility Condition)
    Let $\ell[x]$ be a \sigbv-literal.
    A quantifier-free \sigbv-formula $\ic$
    is an \emph{invertibility condition} for $x$ in $\ell[x]$
    if
    $x \not\in \fv{\ic}$
    and
    $\ic \lequiv \binde{x}{\ell[x]}$ is \tbv-valid.
    \label{def:ic}
  \end{definition}

  An invertibility condition for a literal $\ell[x]$ provides
  the \emph{exact conditions} under which $\ell[x]$ is solvable for $x$.
  We call it an ``invertibility'' condition because we can use Hilbert 
  choice functions to express \emph{all} such conditional solutions
  with a \emph{single} symbolic term, that is, a term whose possible values
  are exactly the solutions for $x$ in $\ell[x]$.
  Recall that a choice function \choice{y}{\varphi[y]} 
  represents a solution for a formula $\varphi[x]$ if there exists one,
  and represents an arbitrary value otherwise.
  We may use a choice function
  to describe inverse solutions
  for a literal $\ell[x]$ 
  with invertibility condition $\ic$
  as
  \choice{y}{(\ic \limpl \ell[y])}.
  For example,
  for the general case of bit-vector multiplication over equality
  the choice function is defined as
  \choice{y}{(\equal{\bvand{(\bvor{\bvneg{s}}{s})}{t}}{t} \;\limpl\;
  \equal{\bvmul{y}{s}}{t})}.

  \begin{lemma}
    \label{lem:ic-choice}
    If $\ic$ is an invertibility condition for an 
    $\choicef$-valid \sigbv-literal $\ell[x]$
    and $r$ is the term $\choice{y}{(\ic \limpl \ell[y])}$,
    then $r$ is $\choicef$-valid and
    $\ell[r] \lequiv \binde{x}{\ell[x]}$
    is \tbv-valid.\begin{conf}\footnote{All proofs can be found in an extended
    version of this paper~\cite{BvQuantReport}.}\end{conf}
  \end{lemma}
  \begin{rep}
  \begin{proof}
    First we show that 
$r = \choice{y}{(\ic \limpl \ell[y])}$ is $\choicef$-valid,
where $\ic$ is an invertibility condition for $y$ in $\ell[y]$. To do so,
since $\ell[y]$ is $\choicef$-valid,
we must show $\binde{y}{\ic \limpl \ell[y]}$ holds in all models of \tbv.
Since $\ic$ is an invertibility condition for $\ell[y]$, 
we have that $y \not\in\fv{\ic}$
and hence this formula is equivalent to $\ic \limpl \binde{y}{\ell[y]}$.
Let $\I$ be any model of \tbv that satisfies $\ic$.
Since $\ic$ is an invertibility condition for $\ell[y]$,
by Definition ~\ref{def:ic}, $\I$ satisfies $\binde{y}{\ell[y]}$ as well.
Thus, $\binde{y}{\ic \limpl \ell[y]}$ holds in all models of \tbv and hence $r$
is $\choicef$-valid.

To show 
$\ell[r] \lequiv \binde{x}{\ell[x]}$ where 
$r$ is $\choice{y}{(\ic \limpl \ell[y])}$,
first consider direction
$\binde{x}{\ell[x]} \limpl
\ell[r]$.
Let $\I$ be any model of \tbv that satisfies $\binde{x}{\ell[x]}$.
By definition of $\choicef$,
$\I$ also satisfies $\ell[\choice{y}{\ell[y]}]$.
Since $\ic$ is an invertibility condition for $\ell[x]$,
from Def.~\ref{def:ic}
we have that
$\ic \lequiv \binde{y}{\ell[y]}$ holds in all models of \tbv,
and thus $\I$ also satisfies $\ic$.
Hence, since $\I$ satisfies $\ell[\choice{y}{\ell[y]}]$, it also 
satifies $\ell[\choice{y}{(\ic \limpl \ell[y])}]$,
which is $\ell[r]$.
Thus,
$\binde{x}{\ell[x]} \limpl
\ell[r]$
is \tbv-valid.
The other direction
$\ell[r] \limpl
\binde{x}{\ell[x]}$
trivially holds in all models of~\tbv.
\qed

  \end{proof}
  \end{rep}

Intuitively, the lemma states that when $\ell[x]$ is satisfiable
(under condition $\ic$),
any value returned by the choice function $\choice{y}{(\ic \limpl \ell[y])}$
is a solution of $\ell[x]$ (and thus $\binde{x}{\ell[x]}$ holds).
Conversely, if there exists a value $v$ for $x$ that makes $\ell[x]$ true, then
there is a model of \tbv that interprets $\choice{y}{(\ic \limpl \ell[y])}$ as $v$.

  Now,
  suppose that \sigbv-literal $\ell$ is again linear in $x$ 
  but that $x$ occurs arbitrarily deep in $\ell$.
  Consider,
  for example,
  a literal
  $\equal{\bvmul{s_1}{(\bvadd{s_2}{x}})}{t}$
  where $x$ does not occur in $s_1$, $s_2$ or $t$.
  We can solve this literal for $x$
  by recursively computing
  the (possibly conditional) inverses of all bit-vector operations
  that involve~$x$.
  That is,
  first
  we solve
  $\equal{\bvmul{s_1}{x'}}{t}$
  for $x'$, where $x'$ is a fresh variable abstracting $\bvadd{s_2}{x}$,
  which yields
  the choice function
  $x' = \choice{y}{(\equal{\bvand{(\bvor{\bvneg{s_1}}{s_1})}{t}}{t}
  \limpl \equal{\bvmul{s_1}{y}}{t})}$.
  Then,
  we solve \equal{\bvadd{s_2}{x}}{x'} for $x$,
  which yields the solution
  $x = \bvsub{x'}{s_2} =
  \bvsub{\choice{y}{(\equal{\bvand{(\bvor{\bvneg{s_1}}{s_1})}{t}}{t}             
   \limpl \equal{\bvmul{s_1}{y}}{t})}}{s_2}$.

  \begin{figure}[t]
    \input{algo/solve.tex}
    \vspace{-4ex}
    \caption{Function $\solvelitf$ for constructing a
    symbolic solution for $x$ given a linear literal $e[x] \rel t$.}
    \label{fig:solve}
  \end{figure}

  Figure~\ref{fig:solve}
  describes in pseudo code the procedure to solve for $x$
  in an arbitrary literal
  $\ell[x] = e[x] \rel t$ that is linear in $x$.
  We assume that $e[x]$ is built over the set of bit-vector operators
  listed in Table~\ref{tab:bvops}.
  Function \solvelitf
  recursively constructs a symbolic solution by computing (conditional)
  inverses
  as follows.
  Let function \getinverse{x}{\ell[x]}
  return a term $t'$ that is the inverse of $x$ in $\ell[x]$,
  i.e., such that $\ell[x] \lequiv \equal{x}{t'}$.
  Furthermore, let function \getsc{x}{\ell[x]}
  return the invertibility condition $\ic$ for $x$ in $\ell[x]$.
  If $e[x]$ has the form $\diamond (e_1, \ldots, e_n)$ with $n > 0$,
  $x$ must occur in exactly one of the subterms $e_1, \ldots, e_n$
  given that $e$ is linear in $x$. 
  Let $d$ be the term obtained from $e$ by replacing $e_i$ (the subterm
  containing $x$)
  with a fresh variable~$x'$.
  We solve for subterm $e_i[x]$ (treating it as a variable~$x'$)
  and compute an inverse \getinverse{x'}{\equal{d[x']}{t}}, if it exists.
  Note that for a disequality $e[x] \tneq t$,
  it suffices to compute the inverse over equality and propagate
  the disequality down.
  (For example,
  for \dist{e_i[x] + s}{t},
  we compute the inverse $t' = \getinverse{x'}{\equal{x'+ s}{t}} = t - s$
  and recurse on $\dist{e_i[x]}{t'}$.)
  If no inverse for $e[x] \rel t$ exists,
  we first determine the invertibility condition $\ic$ for $d[x']$
  via
  \getsc{x'}{d[x'] \rel t}, construct the choice function
  \choice{y}{(\ic \limpl d[y] \rel t)},
  and set it equal to $e_i[x]$, before recursively solving for $x$.
  If $e[x] = x$ and the given literal is an equality,
  we have reached the base case and return $t$ as the solution for $x$.
  Note that in Figure~\ref{fig:solve},
  for simplicity we omitted one case
  for which an inverse can be determined,
  namely
  $x \bvmulf c \teq t$ where $c$ is an odd constant.

  \begin{theorem}
    \label{thm:solve}
    Let $\ell[x]$ be an $\choicef$-valid
    $\Sigma_{BV}$-literal linear in $x$,
    and
    let $r = \solvelit{x,\ell[x]}$.
    Then
    $r$ is $\choicef$-valid,
    $\fv{r} \subseteq \fv{\ell} \setminus \{ x \}$ and
    $\ell[r] \lequiv \binde{x}{\ell[x]}$
    is \tbv-valid.
  \end{theorem}
  \begin{rep}
  \begin{proof}
    We assume without loss of generality 
that $\ell[x]$ is of the form $e[x] \rel t$.
Since $\ell[x]$ is linear with respect to $x$, we have that 
$x \not\in \fv{t}$.
We show all statements of the Theorem 
by structural induction on the term $e[x]$.

Consider the case when $e[x]$ is $x$.
If $\rel$ is $\teq$, then $r$ is $t$.
We have that $r$ is $\choicef$-valid,
$x \not\in \fv{r}$ since $x \not\in \fv{t}$ and 
$t \teq t \lequiv \binde{x}{ x \teq t}$
holds in all models of \tbv.
Otherwise, when $\rel$ is not $\teq$, we have that $r$ is 
$\choice{y}{(\psi \limpl y \rel t)}$ where
$\psi$ is $\getsc{x'}{x' \rel t}$.
By definition of $\getscf$, we have that $\fv{\psi} \subseteq \fv{t}$,
and hence $\fv{r}$ is a subset of $\fv{t}$, 
which is equal to $\fv{ \ell[x] } \setminus \{ x \}$.
Furthermore, since $\psi$ is an invertibility condition for $x'$ in $x' \rel t$,
by Lemma~\ref{lem:ic-choice},
$r$ is $\choicef$-valid and
$r \rel t \lequiv \binde{x}{x \rel t}$.
  
Otherwise, $e[x]$ must be of the form $\diamond (e_1, ..., e_i[x], ..., e_n)$ 
for $n > 0$, where
$x \not\in \fv{e_j}$ for each $i \neq j$.
Let $d[x']$ be 
$\diamond (e_1, ..., e_{i-1}, x', e_{i+1}, ... e_n)$ where
notice that $x \not \in \fv{d[x']}$.
We have that $r$ is $\solvelit{x, e_i[x] \rel_i t_i }$ 
for some relation $\rel_i$ 
and term $t_i$, where $t_i$ is either
$\getinverse{x'}{d[x'] \teq t}$ or 
$\choice{y}{(\getsc{x'}{d[x'] \rel t} \limpl d[y] \rel t)}$.
In both cases, by definition of $\getinversef$ and $\getscf$,
we have that $t_i$ is 
$\choicef$-valid due to Lemma~\ref{lem:ic-choice} and
since $\ell[x]$ is $\choicef$-valid.
Also in both cases, we have that
$\fv{t_i} \subseteq \fv{t} \cup \fv{d[x']} \setminus \{ x' \}
\subseteq \fv{\ell[x]} \setminus \{ x \}$.
Since $e[x] \rel t$ is $\choicef$-valid and linear with 
respect to $x$ and $x \not\in \fv{t_i}$, the literal $e_i[x] \rel_i t_i$
is $\choicef$-valid and linear with respect to $x$ as well.
Thus, by the induction hypothesis, we have that 
$r$ is $\choicef$-valid,
$\fv{r} \subseteq \fv{e_i[x] \rel_i t_i} \setminus \{ x \}$ and the formula
$e_i[r] \rel_i t_i \lequiv \binde{x}{e_i[x] \rel_i t_i}$
holds in all models of \tbv.

Since $\fv{r} \subseteq \fv{e_i[x] \rel_i t_i} \setminus \{ x \}$
and since $\fv{e_i[x]} \subseteq \fv{\ell[x]}$ 
and $\fv{t_i} \subseteq \fv{\ell[x]}$, we have that
$\fv{r} \subseteq \fv{\ell[x]} \setminus \{ x \}$.  

It remains to show 
$e[r] \rel t \lequiv \binde{x}{e[x] \rel t}$.
In the case that $\rel\;\in \{\teq, \tneq\}$
and $\op \in \{ \bvnotf, \bvnegf, \bvaddf \}$, 
we have that 
$\rel_i$ is $\rel$ and 
$t_i$ is $\getinverse{x'}{d[x'] \teq t}$.
By definition of $\getinversef$ and 
since $\rel\;\in \{\teq, \tneq\}$, 
we have that 
$e_i[x] \rel t_i$ and $d[e_i[x]] \rel t$ are equivalent.
Since $d[e_i[x]] = e[x]$, the latter literal is $e[x] \rel t$.
Thus,
since $e_i[r] \rel t_i \lequiv \binde{x}{e_i[x] \rel t_i}$,
we have that 
$e[r] \rel t \lequiv \binde{x}{e[x] \rel t}$.
Otherwise, we have that $\rel_i$ is $\teq$ and $t_i$ is
$\choice{y}{(\getsc{x'}{d[x'] \rel t} \limpl d[y] \rel t)}$.
Since $\psi$ is an invertibility condition for $x'$ in $d[x'] \rel t$,
by Lemma~\ref{lem:ic-choice}, we have that
$d[t_i] \rel t \lequiv \binde{x_i}{d[x_i] \rel t}$
holds in all models of \tbv.
Clearly $e[r] \rel t \limpl \binde{x}{e[x] \rel t}$
holds in all models of \tbv.
Now, consider any model $\I$ of \tbv that satisfies
$\binde{x}{e[x] \rel t}$. 
Since $e[x] = d[e_i[x]]$, we have that $\I$ satisfies 
$\binde{x_i}{d[x_i] \rel t}$ as well.
Thus, since $d[t_i] \rel t \lequiv \binde{x_i}{d[x_i] \rel t}$
holds in all models of \tbv, we have that
$\I$ satisfies $d[t_i] \rel t$.
Notice that $\I$ satisfies $\binde{x}{d[e_i[x]] \rel t}$ by assumption
and $\I$ satisfies $d[t_i] \rel t$.
Thus, since $t_i$ is $\choice{y}{(\psi \limpl d[y] \rel t)}$,  
we have that $\binde{x_i}{e_i[x_i] \teq t_i}$.
Since by the induction hypothesis
$e_i[r] \teq t_i \lequiv \binde{x}{e_i[x] \teq t_i}$
holds in all models of \tbv, we have that
$\I$ satisfies $e_i[r] \teq t_i$.
Thus, since $\I$ satisfies $d[t_i] \rel t$, we have that
$\I$ also satisfies $d[e_i[r]] \rel t$, which is $e[r] \rel t$.
Thus, $e[r] \rel t \Leftarrow \binde{x}{e[x] \rel t}$
holds in all models of \tbv.
\qed

  \end{proof}
  \end{rep}
  

  Tables~\ref{tab:sceq}-\ref{tab:scuineq}
  list the invertibility conditions
  for bit-vector operators
  $\{\bvmulf$, \bvuremf, \bvudivf, \bvandf, \bvorf, \bvlshrf, \bvashrf,
  \bvshlf, $\bvconcatf\}$
  over relations
  $\{$$\teq$, $\tneq$, \bvultf, $\bvugtf\}$.
  Due to space restrictions we omit the conditions for
  signed inequalities since they can be expressed in terms of
  unsigned inequality.
  We omit the invertibility conditions over
  $\{\bvulef$, $\bvugef\}$
  since they 
  can generally be
  constructed by combining the corresponding conditions for equality and
  inequality---although there might be more succinct equivalent conditions.
\begin{conf}
  Finally, we omit the invertibility conditions 
  for operators $\{\bvnotf$, \bvnegf, $\bvaddf\}$
  and
  literals $x \rel t$
  over inequality
  since they are basic bounds checks,
  e.g., for \bvslt{x}{t} we have \dist{t}{\min}.
  The invertibility condition for $x \tneq t$
  and
  for the extract operator
  is \true.\footnote{%
  All the omitted invertibility conditions can be found in the extended
  version of this paper~\cite{BvQuantReport}.
  }
\end{conf}
\begin{rep}

  Table~\ref{tab:scinv} shows the rules for inverse computation for
  bit-vector operators
  $\{\bvnegf$, \bvnotf, \bvaddf, $\bvmulf\}$ over equality.
  Tables~\ref{tab:scgen}-\ref{tab:scsineq2}
  list the remaining invertibility conditions
  for $x \rel t$ and
  bit-vector operators
  $\{\bvnot$, \bvneg, \bvaddf,
  \bvmulf, \bvuremf, \bvudivf, \bvandf, \bvorf, \bvlshrf, \bvashrf,
  \bvshlf, $\bvconcatf\}$.
\end{rep}

  The idea of computing the inverse of bit-vector operators
  has been used successfully in a recent local search approach
  for solving quantifier-free bit-vector constraints
  by Niemetz et al.~\cite{NiemetzPB16}.
  There, target values are propagated
  via inverse value computation.
  In contrast,
  our approach does not determine single inverse values
  based on concrete assignments
  but aims at finding symbolic solutions
  through the generation of conditional inverses.
  In an extended version of that work~\cite{journals/fmsd/NiemetzPB17},
  the same authors present rules for inverse value computation over equality
  but they provide no proof of correctness for them.
  We define invertibility conditions not only over equality but also disequality
  and (un)signed inequality, and verify their correctness
  up to a certain bit-width.

\begin{table}[t]
  \scalebox{1.00}{%
\centering
{%
  \renewcommand{\arraystretch}{1.2}%
  \begin{tabular}{R@{\hskip 2em}L@{\hskip 4em}L}
    \hline\hline

    \ell[x] & \teq & \tneq
    \\ \hline\hline

    \bvmul{x}{s} \rel t
    & \equal{\bvand{(\bvor{\bvneg{s}}{s})}{t}}{t}
    & \boolor{\dist{s}{0}}{\dist{t}{0}}

    \\ \hline

    \bvurem{x}{s} \rel t
    & \bvuge{\bvnot{(\bvneg{s})}}{t}
    & \boolor{\dist{s}{1}}{\dist{t}{0}}
    \\[.5ex]
    \bvurem{s}{x} \rel t
    & \bvuge{\bvand{(\bvsub{\bvadd{t}{t}}{s})}{s}}{t}
    & \boolor{\dist{s}{0}}{\dist{t}{0}}

    \\ \hline

    \bvudiv{x}{s} \rel t
    & \equal{\bvudiv{(\bvmul{s}{t})}{s}}{t}
    & \boolor{\dist{s}{0}}{\dist{t}{\bvnot{0}}}
    \\[1.5ex]
    \bvudiv{s}{x} \rel t
    & \equal{\bvudiv{s}{(\bvudiv{s}{t})}}{t}
    &  \begin{cases}
       \equal{\bvand{s}{t}}{0} & \text{for } \bw{s} = 1\\
       \true & \text{otherwise}
      \end{cases}

    \\[2.5ex] \hline

    \bvand{x}{s} \rel t
    & \equal{\bvand{t}{s}}{t}
    & \boolor{\dist{s}{0}}{\dist{t}{0}}

    \\ \hline

    \bvor{x}{s} \rel t
    & \equal{\bvor{t}{s}}{t}
    & \boolor{\dist{s}{\bvnot{0}}}{\dist{t}{\bvnot{0}}}

    \\ \hline

    \bvlshr{x}{s} \rel t
    & \equal{\bvlshr{(\bvshl{t}{s})}{s}}{t}
    & \boolor{\dist{t}{0}}{\bvult{s}{\bw{s}}}
    \\[.5ex]
    \bvlshr{s}{x} \rel t
    & \bigvee\limits_{i=0}^{\bw{s}} \equal{\bvlshr{s}{i}}{t}
    & \boolor{\dist{s}{0}}{\dist{t}{0}}

    \\[1.5ex] \hline

    \bvashr{x}{s} \rel t
    & (\imp{\bvult{s}{\bw{s}}}{\equal{\bvashr{(\bvshl{t}{s})}{s}}{t}})\;\wedge
    & \true
    \\
    & (\imp{\bvuge{s}{\bw{s}}}{(\boolor{\equal{t}{\bvnot{0}}}{\equal{t}{0}})})
    &
    \\[2ex]
    \bvashr{s}{x} \rel t
    & \multirow[t]{2}{*}{$\bigvee\limits_{i=0}^{\bw{s}} \equal{\bvashr{s}{i}}{t}$}
    & (\boolor{\dist{t}{0}}{\dist{s}{0}})\;\wedge
    \\
    && (\boolor{\dist{t}{\bvnot{0}}}{\dist{s}{\bvnot{0}}})

    \\ \hline

    \bvshl{x}{s} \rel t
    & \equal{\bvshl{(\bvlshr{t}{s})}{s}}{t}
    & \boolor{\dist{t}{0}}{\bvult{s}{\bw{s}}}
    \\[.5ex]
    \bvshl{s}{x} \rel t
    & \bigvee\limits_{i=0}^{\bw{s}} \equal{\bvshl{s}{i}}{t}
    & \boolor{\dist{s}{0}}{\dist{t}{0}}

    \\[1.5ex] \hline

    \bvconcat{x}{s} \rel t
    & \equal{s}{\bvextract{t}{\bw{s}-1}{0}}
    & \true
    \\[.5ex]
    \bvconcat{s}{x} \rel t
    & \equal{s}{\bvextract{t}{\bw{t}-1}{\bw{t}-\bw{s}}}
    & \true

    \\ \hline
  \end{tabular}%
}
}
  \vspace{1ex}
  \caption{Conditions for the invertibility of bit-vector operators
  over (dis)equality.
  Those for\hspace{.1em} \bvmulf, \bvandf and \bvorf are given 
  modulo commutativity of those operators. }
  \label{tab:sceq}
\end{table}

\begin{table}[t]
  \scalebox{1.00}{%
\centering
{%
  \renewcommand{\arraystretch}{1.2}%
  \begin{tabular}{R@{\hskip 2em}L@{\hskip 4em}L}
    \hline\hline

    \ell[x] & \bvultf & \bvugtf
    \\ \hline\hline

    \bvmul{x}{s} \rel t
    & \dist{t}{0}
    & \bvult{t}{\bvor{\bvneg{s}}{s}}

    \\ \hline

    \bvurem{x}{s} \rel t
    & \dist{t}{0}
    & \bvult{t}{\bvnot{(\bvneg{s})}}
    \\[.5ex]
    \bvurem{s}{x} \rel t
    & \dist{t}{0}
    & \bvult{t}{s}

    \\ \hline

    \bvudiv{x}{s} \rel t
    & \booland{\bvult{0}{s}}{\bvult{0}{t}}
    & \bvugt{\bvudiv{\bvnot{0}}{s}}{t}
    \\[.5ex]
    \bvudiv{s}{x} \rel t
    & \booland{\bvult{0}{\bvnot{(\bvand{\bvneg{t}}{s})}}}
              {\bvult{0}{t}}
    & \bvult{t}{\bvnot{0}}

    \\ \hline

    \bvand{x}{s} \rel t
    & \dist{t}{0}
    & \bvult{t}{s}

    \\ \hline

    \bvor{x}{s} \rel t
    & \bvult{s}{t}
    & \bvult{t}{\bvnot{0}}

    \\ \hline

    \bvlshr{x}{s} \rel t
    & \dist{t}{0}
    & \bvult{t}{\bvlshr{\bvnot{s}}{s}}
    \\[.5ex]
    \bvlshr{s}{x} \rel t
    & \dist{t}{0}
    & \bvult{t}{s}

    \\ \hline

    \bvashr{x}{s} \rel t
    & \dist{t}{0}
    & \bvult{t}{\bvnot{0}}
    \\[.5ex]
    \bvashr{s}{x} \rel t
    & \booland{(\boolor{\bvult{s}{t}}{\bvsge{s}{0}})}
              {\dist{t}{0}}
    & \boolor{\bvslt{s}{(\bvlshr{s}{\!\bvnot{t}})}}{\bvult{t}{s}}

    \\ \hline

    \bvshl{x}{s} \rel t
    & \dist{t}{0}
    & \bvult{t}{\bvshl{\bvnot{0}}{s}}
    \\[.5ex]
    \bvshl{s}{x} \rel t
    & \dist{t}{0}
    & \bigvee\limits_{i=0}^{\bw{s}}\bvugt{(\bvshl{s}{i})}{t}

    \\[1.5ex] \hline

    \bvconcat{x}{s} \rel t
    & \imp{\equal{t_x}{0}}
          {\bvult{s}{t_s}}
    & \imp{\equal{t_x}{\bvnot{0}}}{\bvugt{s}{t_s}}
    \\[.1ex]
    & \multicolumn{2}{l}{
      \text{where }
      $t_x = \bvextract{t}{\bw{t}-1}{\bw{t}-\bw{x}}\text{, }
      t_s = \bvextract{t}{\bw{s}-1}{0}$}
    \\[1ex]
    \bvconcat{s}{x} \rel t
    & \booland{\bvule{s}{t_s}}
              {(\imp{\equal{s}{t_s}}{\dist{t_x}{0}})}
    & \booland{\bvuge{s}{t_s}}
              {\imp{\equal{s}{t_s}}{\dist{t_x}{\bvnot{0}}}}
    \\[.1ex]
    & \multicolumn{2}{l}{
      \text{where }
      $t_x = \bvextract{t}{\bw{x}-1}{0}\text{, }
      t_s = \bvextract{t}{\bw{t}-1}{\bw{t}-\bw{s}}$}

    \\ \hline
  \end{tabular}%
}
}
  \vspace{1ex}
  \caption{Conditions for the invertibility of bit-vector operators
  over unsigned inequality.
  Those for\hspace{.1em} \bvmulf, \bvandf and \bvorf are given 
  modulo commutativity of those operators. }
  \label{tab:scuineq}
\end{table}

\begin{rep}
\begin{table}

\centering
{%
  \renewcommand{\arraystretch}{1.2}%
  \begin{tabular}{R@{\hskip 2em}L@{\hskip 2em}L@{\hskip 2em}L@{\hskip 2em}L@{\hskip 0.5em}l}
    \hline\hline
    \ell[x]
    & \bvneg{x} \teq t
    & \bvnot{x} \teq t
    & \bvadd{x}{s} \teq t
    & \bvmul{x}{s} \teq t \hspace{.25em}
    & with $s$ const
    \\ \hline\hline
    \getinverse{x}{\ell[x]}
    & \bvneg{t}
    & \bvnot{t}
    & \bvsub{t}{s}
    & \bvmul{t}{s^{-1}}\hspace{.25em}
    & with $\bvmul{s}{s^{-1}} \teq 1$
    \\ \hline
  \end{tabular}%
}

  \vspace{2ex}
  \caption{Inverse computation for bit-vector operators 
  $\{\bvnegf$, \bvnotf, \bvaddf, $\bvmulf\}$ over $\teq$.
  Those for\hspace{.1em} \bvaddf and \bvmulf are given 
  modulo commutativity of those operators. }
  \label{tab:scinv}
\end{table}

\begin{table}

\centering
{%
  \renewcommand{\arraystretch}{1.2}%
  \begin{tabular}{R@{\hskip 2.5em}C@{\hskip 2.5em}C@{\hskip 2.5em}C@{\hskip 2.5em}C@{\hskip 2.5em}C}
    \hline\hline

    \ell[x] & \bvultf & \bvugtf  & \bvsltf & \bvsgtf & \bvulef, \bvugef, \bvslef, \bvsgef
    \\ \hline\hline

    x \rel t
    & \multirow{4}{*}{\dist{t}{0}}
    & \multirow{4}{*}{\dist{t}{\bvnot{0}}}
    & \multirow{4}{*}{\dist{t}{\mins}}
    & \multirow{4}{*}{\dist{t}{\maxs}}
    
    & \multirow{4}{*}{\true}

    \\
    \bvneg{x} \rel t
    \\
    \bvnot{x} \rel t
    \\
    \bvadd{x}{s} \rel t
    \\ \hline
  \end{tabular}%
}

  \vspace{2ex}
  \caption{Conditions for the invertibility for $x \rel t$ and
  bit-vector operators
  $\{\bvnegf$, \bvnotf,  $\bvaddf\}$
  over inequality.
  The one for \bvaddf given 
  modulo commutativity of \bvaddf. }
  \label{tab:scgen}
\end{table}

\begin{table}
  \centering
{%
  \renewcommand{\arraystretch}{1.2}%
  \begin{tabular}{R@{\hskip 2em}L@{\hskip 2em}L}
    \hline\hline

    \ell[x] & \bvulef & \bvugef
    \\ \hline\hline

    \bvmul{x}{s} \rel t
    & \true
    & \bvuge{\bvor{\bvneg{s}}{s}}{t}

    \\ \hline
    \bvurem{x}{s} \rel t
    & \true
    & \bvuge{\bvnot{(\bvneg{s})}}{t}
    \\[.5ex]
    \bvurem{s}{x} \rel t
    & \true
    & \boolor{\bvuge{\bvand{(\bvsub{\bvadd{t}{t}}{s})}{s}}{t}}{\bvult{t}{s}}

    \\ \hline

    \bvudiv{x}{s} \rel t
    & \bvuge{\bvor{s}{t}}{\;\bvnot{(\bvneg{s})}}
    & \equal{\bvand{\bvudiv{(\bvmul{s}{t})}{t}}{s}}{s}
    \\[.5ex]
    \bvudiv{s}{x} \rel t
    & \bvult{0}{\bvor{\;\bvnot{s}}{t}}
    & \true

    \\ \hline

    \bvand{x}{s} \rel t
    & \true
    & \bvuge{s}{t}

    \\ \hline

    \bvor{x}{s} \rel t
    & \bvuge{t}{s}
    & \true

    \\ \hline
    \bvlshr{x}{s} \rel t
    & \true
    & \equal{\bvlshr{(\bvshl{t}{s})}{s}}{t}
    \\[.5ex]
    \bvlshr{s}{x} \rel t
    & \true
    & \bvuge{s}{t}

    \\ \hline

    \bvashr{x}{s} \rel t
    & \true
    & \true
    \\[.5ex]
    \bvashr{s}{x} \rel t
    & \boolor{\bvult{s}{\mins}}{\bvuge{t}{s}}
    & \boolor{\bvuge{s}{\;\bvnot{s}}}{\bvuge{s}{t}}

    \\ \hline

    \bvshl{x}{s} \rel t
    & \true
    & \bvuge{\bvshl{\bvnot{0}}{s}}{t}
    \\[.5ex]
    \bvshl{s}{x} \rel t
    & \true
    & \bigvee\limits_{i=0}^{\bw{s}}\bvuge{(\bvshl{s}{i})}{t}

    \\[1.5ex] \hline

    \bvconcat{x}{s} \rel t
    & \imp{\equal{t_x}{0}}{\bvule{s}{t_s}}
    & \imp{\equal{t_x}{\;\bvnot{0}}}{\bvuge{s}{t_s}}
    \\[.1ex]
    & \multicolumn{2}{l}{
      \text{where }
      $t_x = \bvextract{t}{\bw{t}-1}{\bw{t}-\bw{x}}\text{, }
      t_s = \bvextract{t}{\bw{s}-1}{0}$}
    \\[1ex]
    \bvconcat{s}{x} \rel t
    & \bvule{s}{t_s}
    & \bvuge{s}{t_s}
    \\[.1ex]
    & \multicolumn{2}{l}{
      \text{where }
      $t_x = \bvextract{t}{\bw{x}-1}{0}\text{, }
      t_s = \bvextract{t}{\bw{t}-1}{\bw{t}-\bw{s}}$}

    \\ \hline
  \end{tabular}%
}

  \vspace{2ex}
  \caption{Conditions for the invertibility of bit-vector operators
  over \bvulef and \bvugef.
  Those for\hspace{.1em} \bvmulf, \bvandf and \bvorf are given 
  modulo commutativity of those operators. }
  \label{tab:scuineq2}
\end{table}
\begin{table}
  \centering
{%
  \renewcommand{\arraystretch}{1.2}%
  \begin{tabular}{R@{\hskip 1.4em}L@{\hskip 1.0em}L}
    \hline\hline

    \ell[x] & \bvsltf & \bvsgtf
    \\ \hline\hline

    \bvmul{x}{s} \rel t
    & \bvslt{\bvand{\bvnot{(\bvneg{t})}}{(\bvor{\bvneg{s}}{s})}}{t}
    & \bvslt{t}{\bvsub{t}{(\bvor{(\bvor{s}{t})}{\bvneg{s}})}}

    \\ \hline
    \bvurem{x}{s} \rel t
    & \bvslt{\bvnot{t}}{(\bvor{\bvneg{s}}{\bvneg{t}})}
    & \imp{(\bvsgt{s}{0}}{\bvslt{t}{\;\bvnot{(\bvneg{s})}}})\;\wedge
    \\
    && \imp{(\bvsle{s}{0}}{\dist{t}{\maxs}})\;\wedge
    \\
    && \boolor{(\dist{t}{0}}{\dist{s}{1}})
    \\[1.5ex]
    \bvurem{s}{x} \rel t
    & \boolor{\bvslt{s}{t}}{\bvslt{0}{t}}

    & \imp{(\bvsge{s}{0}}{\bvsgt{s}{t}})\;\wedge
    \\
    && \imp{(\bvslt{s}{0}}{\bvsgt{(\bvlshr{(\bvsub{s}{1})}{1})}{t}})

    \\ \hline

    \bvudiv{x}{s} \rel t
    & \imp{\bvsle{t}{0}}{\bvslt{\bvudiv{\mins}{s}}{t}}
    & \boolor{\bvsgt{\bvudiv{\bvnot{0}}{s}}{t}}
             {\bvsgt{\bvudiv{\maxs}{s}}{t}}
    \\[.5ex]
    \bvudiv{s}{x} \rel t
    & \boolor{\bvslt{s}{t}}{\bvsge{t}{0}}
    %
    & \begin{cases}
        \bvsgt{s}{t} & \text{for \bw{s} = 1} \\[1ex]
        \imp{(\bvsge{s}{0}}{\bvsgt{s}{t})}\;\wedge & \text{otherwise} \\
        \imp{(\bvslt{s}{0}}{\bvsgt{\bvlshr{s}{1}}{t})} &
      \end{cases}

    \\[5ex] \hline

    \bvand{x}{s} \rel t
    & \bvslt{\bvand{\bvnot{(\bvneg{t})}}{s}}{t}
    & \bvslt{t}{\bvand{s}{\maxs}}

    \\ \hline

    \bvor{x}{s} \rel t
    & \bvslt{\bvor{\bvnot{(\bvsub{s}{t})}}{s}}{t}
    & \bvslt{t}{(\bvor{s}{\maxs})}
    \\
    \bvor{s}{x} \rel t & &

    \\ \hline
    \bvlshr{x}{s} \rel t
    & \bvslt{\bvlshr{\bvnot{(\bvneg{t})}}{s}}{t}
    & \bvslt{t}{\bvlshr{(\bvshl{\maxs}{s})}{s}}
    \\[.5ex]
    \bvlshr{s}{x} \rel t
    & \boolor{\bvslt{s}{t}}{\bvslt{0}{t}}
    & (\imp{\bvslt{s}{0}}{\bvsgt{\bvlshr{s}{1}}{t}})\;\wedge
    \\
    && (\imp{\bvsge{s}{0}}{\bvsgt{s}{t}})

    \\ \hline

    \bvashr{x}{s} \rel t
    & \bvslt{\bvashr{\mins}{s}}{t}
    & \bvslt{t}{\bvlshr{\maxs}{s}}
    \\[.5ex]
    \bvashr{s}{x} \rel t
    & \boolor{\bvslt{s}{t}}{\bvslt{0}{t}}
    & \booland{\bvslt{t}{\bvand{s}{\maxs}}}{\bvslt{t}{\bvor{s}{\maxs}}}

    \\ \hline

    \bvshl{x}{s} \rel t
    & \bvslt{\bvshl{(\bvlshr{\mins}{s})}{s}}{t}
    & \bvslt{t}{\bvand{(\bvshl{\maxs}{s})}{\maxs}}
    \\[.5ex]
    \bvshl{s}{x} \rel t
    & \bvult{\bvshl{\mins}{s}}{\bvadd{t}{\mins}}
    & \bigvee\limits_{i=0}^{\bw{s}}\bvsgt{(\bvshl{s}{i})}{t}

    \\[1.5ex] \hline

    \bvconcat{x}{s} \rel t
    & \imp{\equal{t_x}{\mins}}{\bvult{s}{t_s}}
    & \imp{\equal{t_x}{\maxs}}{\bvugt{s}{t_s}}
    \\[.1ex]
    & \multicolumn{2}{l}{%
      \text{where }
      $t_x = \bvextract{t}{\bw{t}-1}{\bw{t}-\bw{x}}\text{, }
      t_s = \bvextract{t}{\bw{s}-1}{0}$}
    \\[1ex]
    \bvconcat{s}{x} \rel t
    & \booland{(\bvsle{s}{t_s})}
              {(\imp{\equal{s}{t_s}}{\dist{t_x}{0}})}
    & \booland{(\bvsge{s}{t_s})}
              {(\imp{\equal{s}{t_s}}{\dist{t_x}{\;\bvnot{0}}})}
    \\[.1ex]
    & \multicolumn{2}{l}{%
      \text{where }
      $t_x = \bvextract{t}{\bw{x}-1}{0}\text{, }
      t_s = \bvextract{t}{\bw{t}-1}{\bw{t}-\bw{s}}$}

    \\ \hline
  \end{tabular}%
}

  \vspace{1ex}
  \caption{Conditions for the invertibility of bit-vector operators
  over \bvsltf and \bvsgtf.
  Those for\hspace{.1em} \bvmulf, \bvandf and \bvorf are given 
  modulo commutativity of those operators. }
  \label{tab:scsineq}
\end{table}

\begin{table}
  \centering
{%
  \renewcommand{\arraystretch}{1.2}%
  \begin{tabular}{R@{\hskip 2em}L@{\hskip 2em}L}
    \hline\hline

    \ell[x] & \bvslef & \bvsgef
    \\ \hline\hline

    \bvmul{x}{s} \rel t
    & \bvnot{(\booland{\equal{s}{0}}{\bvslt{t}{s}})}
    & \bvsge{\bvand{(\bvor{\bvneg{s}}{s})}{\maxs}}{t}

    \\ \hline
    \bvurem{x}{s} \rel t
    & \bvslt{\bvnot{0}}{\bvand{\bvneg{s}}{t}}
    & \boolor{\bvslt{t}{s}}{\bvsge{0}{s}}
    \\[.5ex]
    \bvurem{s}{x} \rel t
    & \boolor{\bvult{t}{\mins}}{\bvsge{t}{s}}
    & (\imp{\bvsge{s}{0}}{\bvsge{s}{t}})\;\wedge
    \\
    && (\imp{\booland{(\bvslt{s}{0}}{\bvsge{t}{0}})}
                     {\bvugt{\bvsub{s}{t}}{t}})

    \\ \hline

    \bvudiv{x}{s} \rel t
    & (\equal{\bvudiv{(\bvmul{s}{t})}{s}}{t})\;\vee
    & \boolor{(\bvsge{\bvudiv{\bvnot{0}}{s}}{t})}
             {(\bvsge{\bvudiv{\maxs}{s}}{t})}
    \\
    & (\imp{\bvsle{t}{0}}{\bvslt{\bvudiv{\mins}{s}}{t}})
    \\[.5ex]
    \bvudiv{s}{x} \rel t
    & \boolor{\bvsge{t}{\;\bvnot{0}}}{\bvsge{t}{s}}
    & (\imp{\bvsge{s}{0}}{\bvsge{s}{t}})\;\wedge
    \\
    && (\imp{\bvslt{s}{0}}{\bvsge{\bvlshr{s}{1}}{t}})
    \\ \hline

    \bvand{x}{s} \rel t
    & \bvuge{s}{\bvand{t}{\mins}}
    & \boolor{\equal{\bvand{s}{t}}{t}}{\bvslt{t}{\bvand{(\bvsub{t}{s})}{s}}}

    \\ \hline

    \bvor{x}{s} \rel t
    & \bvsge{t}{\bvor{s}{\mins}}
    & \bvand{s}{t}

    \\ \hline
    \bvlshr{x}{s} \rel t
    & \bvsge{t}{\bvlshr{t}{s}}
    & \imp{\dist{s}{0}}{\bvsge{\bvlshr{\;\bvnot{0}}{s}}{t}}
    \\[.5ex]
    \bvlshr{s}{x} \rel t
    & \boolor{\bvult{t}{\mins}}{\bvsge{t}{s}}
    & (\imp{\bvslt{s}{0}}{\bvsge{\bvlshr{s}{1}}{t}})\;\wedge
    \\
    && (\imp{\bvsge{s}{0}}{\bvsge{s}{t}})

    \\ \hline

    \bvashr{x}{s} \rel t
    & \bvsge{t}{\;\bvnot{(\bvlshr{\maxs}{s})}}
    & \bvsge{\bvlshr{\maxs}{s}}{t}
    \\[.5ex]
    \bvashr{s}{x} \rel t
    & \boolor{\bvsge{t}{0}}{\bvsge{t}{s}}
    & \boolor{\bvuge{t}{\;\bvnot{t}}}{\bvsge{s}{t}}

    \\ \hline

    \bvshl{x}{s} \rel t
    & \bvult{\bvlshr{t}{(\bvlshr{t}{s})}}{\mins}
    & \bvsge{\bvand{(\bvshl{\maxs}{s})}{\maxs}}{t}
    \\[.5ex]
    \bvshl{s}{x} \rel t
    & \bvult{\bvlshr{t}{s}}{\mins}
    & \bigvee\limits_{i=0}^{\bw{s}}\bvsge{(\bvshl{s}{i})}{t}

    \\[1.5ex] \hline

    \bvconcat{x}{s} \rel t
    & \imp{\equal{t_x}{\mins}}{\bvule{s}{t_s}}
    & \imp{\equal{t_x}{\maxs}}{\bvuge{s}{t_s}}
    \\[.1ex]
    & \multicolumn{2}{l}{
      \text{where }
      $t_x = \bvextract{t}{\bw{t}-1}{\bw{t}-\bw{x}}\text{, }
      t_s = \bvextract{t}{\bw{s}-1}{0}$}
    \\[1ex]
    \bvconcat{s}{x} \rel t
    & \bvsle{s}{t_s}
    & \bvsge{s}{t_s}
    \\[.1ex]
    & \multicolumn{2}{l}{
      \text{where }
      $t_x = \bvextract{t}{\bw{x}-1}{0}\text{, }
      t_s = \bvextract{t}{\bw{t}-1}{\bw{t}-\bw{s}}$}

    \\ \hline
  \end{tabular}%
}

  \vspace{1ex}
  \caption{Conditions for the invertibility of bit-vector operators
  over \bvslef and \bvsgef.
  Those for\hspace{.1em} \bvmulf, \bvandf and \bvorf are given 
  modulo commutativity of those operators. }
  \label{tab:scsineq2}
\end{table}
\end{rep}

\subsection{Synthesizing Invertibility Conditions}
  %
  %
  %

  We have defined invertibility conditions for all bit-vector operators
  in $\sig_{BV}$ where no general inverse exists (162 in total).
  A noteworthy aspect of this work is that we were able to leverage
  syntax-guided synthesis (SyGuS) technology~\cite{AlurBJMRSSSTU13} 
  to help identify these conditions.
  The problem of finding invertibility conditions for a literal of the form
  $x \op s \rel t$ (or, dually, $s \op x \rel t$) linear in $x$
  can be recast 
  as a SyGuS problem
  by asking whether there exists
  a binary Boolean function $C$ such that the (second-order) formula
  $\binde{C\forall s\forall t}{((
   \binde{x}{x \op s \rel t )\lequiv C(s,t))}}$
  is satisfiable.
  If a SyGuS solver is able to synthesize the function $C$, 
  then $C$ can be used as the invertibility condition for $x \op s \rel t$.
  To simplify the SyGuS problem
  we chose a bit-width of 4 for $x$, $s$, and $t$
  and eliminated the quantification over $x$ in the formula above by 
  by expanding it to
  \vspace{-1ex}
  \begin{equation*}
    \binde{C\forall s\forall t}{
      (\bigvee\limits_{i=0}^{15} i \op s \rel t ) \lequiv C(s,t)}
  \end{equation*}
  Since the search space for SyGuS solvers heavily depends on 
  the input grammar (which defines the solution space for $C$), 
  we decided to use two grammars with the same set 
  of Boolean connectives but different sets of bit-vector operators:
  \begin{eqnarray*}
    O_r & = &
           \{ \neg, \land, \teq, \bvultf, \bvsltf, 0, \mins, \maxs, s, t,
              \bvnotf, \bvnegf, \bvandf, \bvorf \}
    \\
    O_g & = &
           \{ \neg, \land, \lor, \teq, \bvultf, \bvsltf, \bvugef, \bvsgef,
              0, \mins, \maxs, s, t,
              \bvnotf, \bvaddf, \bvnegf, \bvandf, \bvorf, \bvlshrf, \bvshlf \}
  \end{eqnarray*}
  The selection of 
  constants in the grammar
  turned out to be crucial for finding solutions, e.g.,
  by adding \mins and \maxs we were able to synthesize
  substantially more invertibility conditions for signed inequalities.
  For each of the two sets of operators,
  we generated 140 SyGuS problems\footnote{
  Available at \url{https://cvc4.cs.stanford.edu/papers/CAV2018-QBV/}}, 
  one for each combination of
  bit-vector operator $\op \in$
  \{\bvmulf, \bvuremf, \bvudivf, \bvandf, \bvorf, \bvlshrf, \bvashrf,
    $\bvshlf\}$
  over relation $\rel\;\in$
  \{\teq, \tneq,
    \bvultf, \bvulef, \bvugtf, \bvugef,
    \bvsltf, \bvslef, \bvsgtf, $\bvsgef\}$,
  and used the SyGuS extension of the
  \cvc solver~\cite{ReynoldsDKTB15} to solve these problems.

  Using operators $O_r$ ($O_g$) we were able to synthesize 98 (116) out of 140
  invertibility conditions, with 118 unique solutions overall.
  When we found more than one solution
  for a condition
  (either with operators $O_r$ and $O_g$, or manually)
  we chose the one that involved the smallest number of bit-vector operators.
  Thus, we ended up using 79 out of 118 synthesized conditions
  and 83 manually crafted conditions.

  In some cases, the SyGuS approach was able to synthesize invertibility
  conditions that were smaller than those we had manually crafted.
  For example,
  we manually defined the invertibility condition for $\bvmul{x}{s} \teq t$
  as
  $\boolor{(t \teq 0)}{(\booland{
    \bvuge{(\bvand{t}{\bvneg{t}})}{(\bvand{s}{\bvneg{s}})}}{(s \tneq 0)})}$.
  With SyGuS we obtained
  $(\bvand{(\bvor{\bvneg{s}}{s})}{t}) \teq t$.
  For some other cases, however,
  the synthesized solution involved more bit-vector operators than needed.
  For example,
  for ${\bvurem{x}{s} \tneq t}$ we manually defined the invertibility
  condition ${\boolor{(s \tneq 1)}{(t \tneq 0)}}$, whereas SyGuS produced the
  solution $\bvor{\bvnot{(\bvneg{s})}}{t} \tneq 0$.
  For the majority of invertibility conditions, finding a solution did not
  require more than one hour of CPU time on an Intel Xeon E5-2637
  with~3.5GHz.
  Interestingly, the most time-consuming synthesis task
  (over 107 hours of CPU time) was finding 
  condition
  $\bvuge{\bvand{(\bvsub{(\bvadd{t}{t})}{s})}{s}}{t}$
  for
  ${\equal{\bvurem{s}{x}}{t}}$.
  A small number of synthesized solutions were only correct for a bit-width
  of 4, e.g,
  solution
  $\bvslt{\bvshl{(\bvshl{\bvnot{s}}{s})}{s}}{t}$
  for
  $\bvslt{\bvudiv{x}{s}}{t}$.
  In total, we found 6 width-dependent synthesized solutions,
  all of them for bit-vector operators \bvudivf and \bvuremf.
  For those, we used the manually crafted invertibility conditions instead.

\subsection{Verifying Invertibility Conditions}
  We verified the correctness of all 162 invertibility conditions
  for bit-widths from~1 to~65 by checking
  for each bit-width the \tbv-unsatisfiability of the formula
  $\boolnot{(\ic \lequiv \binde{x}{\ell[x]})}$
  where $\ell$ ranges over the literals in Tables~\ref{tab:sceq}--\ref{tab:scuineq}
  with $s$ and $t$ replaced by fresh constants,
  and $\ic$ is the corresponding invertibility condition.
  
  In total, we generated 12,980 verification problems and used all
  participating solvers 
  of the quantified bit-vector division of
  SMT-competition 2017 to verify them.
  For each solver/benchmark pair we used a CPU time limit of one hour and a
  memory limit of 8GB on the same machines as those mentioned in the previous section.
  We consider an invertibility condition to be verified for a certain bit-width
  if at least one of the solvers was able to report unsatisfiable 
  for the corresponding formula within the given time limit.
  Out of the 12,980 instances,
  we were able to verify 12,277 (94.6\%).

  Overall, all verification tasks (including timeouts)
  required a total of 275 days of CPU time.
  The success rate of each individual solver was
  91.4\% for \boolector,
  85.0\% for \cvc,
  50.8\% for \qiiib,
  and 92\% for \ziii.
  We observed that on 30.6\% of the problems, \qiiib exited with a Python
  exception without returning any result.
  For bit-vector operators
  \{\bvnotf, \bvnegf, \bvaddf, \bvandf, \bvorf, \bvlshrf, \bvashrf, \bvshlf,
    $\bvconcatf\}$,
  over all relations,
  and for operators
  \{\bvmulf, \bvudivf, $\bvuremf\}$
  over relations $\{\tneq, \bvulef, \bvslef\}$,
  we were able to verify all invertibility conditions 
  for all bit-widths in the range 1--65.
  Interestingly, no solver was able to verify the invertibility conditions
  for $\bvslt{\bvurem{x}{s}}{t}$ with a bit-width of 54 and
  $\bvult{\bvurem{s}{x}}{t}$ with bit-widths 35-37
  within the allotted time.
  We attribute this to the underlying heuristics used by the SAT solvers
  in these systems.
  All other conditions for \bvsltf and \bvultf were verified for all
  bit-vector operators up to bit-width 65.
  The remaining conditions for operators
  \{\bvmulf, \bvudivf, $\bvuremf\}$
  over relations
  \{\teq, \bvugtf, \bvugef, \bvsgtf, $\bvsgef\}$
  were verified up to at least a bit-width of 14.
  We discovered 3 conditions for
  $\bvudiv{s}{x} \rel t$ with $\rel\;\in \{\tneq, \bvsgtf, \bvsgef\}$
  that were not correct for a bit-width of 1.
  For each of these cases,
  we added an additional invertibility condition that correctly
  handles that case.
 
  We leave to future work the task of formally proving that our 
  invertibility conditions 
  are correct for all bit-widths.
  Since this will most likely require the development of an interactive proof,
  we could leverage some recent work by Ekici et al.~\cite{EkiEtAl-CAV-17} 
  that includes a formalization in the Coq proof assistant of the SMT-LIB theory 
  of bit-vectors.

\section{Counterexample-Guided Instantiation for Bit-Vectors}
\label{sec:cegqibv}

In this section, we leverage techniques from the previous section
for constructing symbolic solutions to bit-vector constraints
to define a novel instantiation-based technique 
for quantified bit-vector formulas.
\begin{rep}
At a high level, we use a \emph{counterexample-guided} approach
for quantifier instantiation that adds new instances
to a set of quantifier-free clauses based on models
for the negated input formula.
The procedure terminates
if it finds a set of instances that is unsatisfiable
or entails the negation of the input formula.
\end{rep}%
\begin{conf}
We first briefly present the overall theory-independ\-ent
procedure we use for quantifier instantiation
and then show how it can be specialized to quantified bit-vectors using
invertibility conditions.
\end{conf}

\begin{figure}[t]
  \input{algo/cegqis.tex}
  \vspace{-4ex}
  \caption{A counterexample-guided quantifier instantiation
  procedure $\cegqi{\selfun}$, parameterized by a selection function~$\selfun$,
  for determining the $T$-satisfiability of
  $\binde{\vec{y}\forall\vec{x}}{\psi[\vec{y}, \vec{x}]}$ 
  with $\psi$ quantifier-free and $\fv{\psi} = \vec{y} \cup \vec{x}$.
  }
  \label{fig:proc-qi}
\end{figure}

We use a counterexample-guided approach for quantifier instantiation,
as given by procedure $\cegqi{\selfun}$ in Figure~\ref{fig:proc-qi}.
To simplify the exposition here, we focus on input problems expressed 
as a single formula in prenex normal form and with up 
to one quantifier alternation. 
We stress, though, that the approach applies in general to arbitrary
sets of quantified formulas in some $\Sigma$-theory $T$ 
with a decidable quantifier-free fragment.
The procedure checks via instantiation the $T$-satisfiability 
of a quantified input formula $\varphi$ of the form 
$\binde{\vec{y}\forall\vec{x}}{\psi[\vec{x},\vec{y}]}$
where $\psi$ is quantifier-free and $\vec{x}$ and $\vec{y}$ 
are possibly empty sequences of variables.
It maintains an evolving set $\Gamma$, initially empty,
of quantifier-free instances of the input formula.
During each iteration of the procedure's loop,
there are three possible cases:
1) if $\Gamma$ is $T$-unsatisfiable, 
the input formula $\varphi$ is also $T$-unsatisfiable and ``unsat'' is returned;
2)  if $\Gamma$ is $T$-satisfiable but not together with
with $\lnot \psi[\vec{y}, \vec{x}]$,
the negated body of $\varphi$,
then $\Gamma$ entails $\varphi$ in $T$, 
hence $\varphi$ is $T$-satisfiable and ``sat'' is returned.
3) If neither of previous cases holds, 
the procedure adds to $\Gamma$ an instance of $\psi$ obtained by replacing
the variables \vec{x} with some terms \vec{t},
and continues.
The procedure $\cegqi{}$ is parametrized by a \emph{selection function} $\selfun$
that generates the terms \vec{t}.

\begin{definition}
\label{def:selfun} (Selection Function)
A \emph{selection function}
takes as input a tuple of variables $\vec{x}$,
a model $\I$ of $T$, 
a quantifier-free $\Sigma$-formula $\psi[\vec x]$,
and a set $\Gamma$ of $\Sigma$-formulas such that
$\vec{x} \not \in \fv{\Gamma}$ and
$\I \models \Gamma \cup \{ \lnot \psi \}$.
It returns a tuple of $\choicef$-valid terms \vec{t} of the same type as \vec{x} 
such that $\fv{\vec{t}} \subseteq \fv{\psi} \setminus \vec{x}$.
\end{definition}

\begin{definition}
\label{def:selfun-prop}
Let $\psi[\vec x]$ be a quantifier-free $\Sigma$-formula.
A selection function is:
\begin{enumerate}[topsep=1ex]
\item 
\emph{Finite for $\vec x$ and $\psi$} 
if there is a finite set $\mathcal{S}^\ast$ such that
$\selfun( \vec{x}, \psi, \I, \Gamma ) \in \mathcal{S}^\ast$
for all legal inputs $\I$ and $\Gamma$.
\item 
\emph{Monotonic for $\vec x$ and $\psi$} 
if for all legal inputs $\I$ and $\Gamma$,
$\selfun(\vec{x}, \psi, \I, \Gamma ) = \vec{t}$ 
only if $\psi[ \vec{t} ] \not\in \Gamma$.
\end{enumerate}
\end{definition}

Procedure $\cegqi{\selfun}$ is refutation-sound and model-sound
for any selection function $\selfun$, and terminating 
for selection functions that are finite and monotonic.

\begin{theorem}[Correctness of $\cegqi{\selfun}$]
\label{thm:cegqi}
Let $\selfun$ be a selection function and 
let $\varphi = \binde{\vec{y}\forall\vec{x}}{\psi[ \vec{y}, \vec{x}]}$
be a legal input for $\cegqi{\selfun}$.
Then the following hold.
\begin{enumerate}[topsep=1ex]
\item If $\cegqi{\selfun}( \varphi )$ returns ``unsat'', 
then $\varphi$ is $T$-unsatisfiable.
\item If $\cegqi{\selfun}( \varphi )$ returns ``sat''
for some final $\Gamma$, 
then $\varphi$ is $T$-equivalent to 
$\binde{\vec{y}}{\bigwedge_{\gamma \in \Gamma} \gamma}$.
\item If $\selfun$ is finite and monotonic for $\vec x$ and $\psi$, 
then $\cegqi{\selfun}( \varphi )$ terminates.
\end{enumerate}
\end{theorem}
\begin{rep}
\begin{proof}
  We show each part of the theorem below.
Note that by the definition of $\cegqi{\selfun}$ and 
since $\selfun$ is a selection function, all inputs $\I$ and $\Gamma$ 
given to $\selfun$ in the loop of this function are legal inputs.
Also note that for the first two parts, 
we have that $\cegqi{\selfun}( \varphi )$ terminates
in a state where $\Gamma$
is a set of formulas of the form $\psi[ \vec{y}, \vec{t} ]$
where, since $\selfun$ is a selection function,
\vec{t} is a tuple of $\choicef$-valid terms and 
$\fv{\vec{t}} \subseteq \vec{y}$.

Part 1) 
By definition of $\cegqi{\selfun}$,
if $\cegqi{\selfun}( \varphi )$ returns ``unsat'',
then $\Gamma$ is $T$-unsatisfiable.
By the definition of $\cegqi{\selfun}$,
we have that $\Gamma$ is a set of formulas 
of the form $\psi[ \vec{y}, \vec{t} ]$.
For each $\psi[ \vec{y}, \vec{t} ] \in \Gamma$,
we have that \vec{t} is $\choicef$-valid since $\selfun$ 
is a selection function, and hence
$\forall \vec{x} \psi[ \vec{y}, \vec{x}] \limpl \psi[ \vec{y}, \vec{t} ]$
holds in all models of $T$.
Since $\Gamma$ is $T$-unsatisfiable, 
$\forall \vec{x} \psi[ \vec{y}, \vec{x}]$ is $T$-unsatisfi\-able,
and hence $\varphi$ is as well.

Part 2) 
By definition of $\cegqi{\selfun}$,
if $\cegqi{\selfun}( \varphi )$ returns ``sat'', the
$\Gamma$ is $T$-satisfiable and 
$\Gamma' = \Gamma \cup \{ \lnot \psi[\vec{y}, \vec{x}] \}$ is
$T$-unsatisfiable.
For each $\psi[ \vec{y}, \vec{t} ] \in \Gamma$,
we have that $\fv{\vec{t}} \subseteq \vec{y}$ and hence
$\fv{\Gamma} \subseteq \vec{y}$.
Since $\vec{x} \not\in \fv{\Gamma}$, we have that
$\Gamma \cup \{ \binde{\vec{x}}{ \neg \psi[\vec{y}, \vec{x}] } \}$
is also $T$-unsatisfiable.
Let $\I$ be an arbitrary model of $\Gamma$.
Since $\Gamma'$ is unsatisfiable,
it must be that
$\I \not\models \binde{\vec{x}}{ \neg \psi[\vec{y}, \vec{x}] }$
and hence $\I \models \binda{\vec{x}}{ \psi[\vec{y}, \vec{x}] }$.
Thus 
$\Gamma \limpl \binde{\vec{x}}{ \neg \psi[\vec{y}, \vec{x}] }$ and hence
$\binde{\vec{y}}{\Gamma} \limpl \varphi$ holds in all models of $T$.
By the same reasoning as Part 1,
we have that 
$\varphi \limpl \binde{\vec{y}}{\Gamma}$ holds in all models as well.
Thus, $\varphi$ is equivalent to $\binde{\vec{y}}{\Gamma}$.

Part 3) 
Assume $\selfun$ is monotonic and finite for $\psi[ \vec{y}, \vec{x} ]$.
Since it is finite, let $\selfun^\ast$ be a finite set such that 
$\selfun( \vec{x}, \psi, \I, \Gamma ) \in \mathcal{S}^\ast$ for all valid 
inputs $\I, \Gamma$.
Since it is monotonic, each iteration of the loop adds a new formula
from $\selfun^\ast$ to $\Gamma$. Since $\selfun^\ast$ is finite, 
the number of iterations of this loop is bounded by the size of $\selfun^\ast$.
Hence, $\cegqi{\selfun}( \varphi )$ terminates.
\qed

\end{proof}
\end{rep}

\noindent
Thanks to this theorem, 
to define a $T$-satisfiability procedure for quantified $\Sigma$-formulas, 
it suffices to define a selection function satisfying the criteria 
of Definition~\ref{def:selfun}.
We do that in the following section for \tbv.

\begin{figure}[t]

\begin{framed}
  \setlist{nolistsep}
  $\selfunbv{c}( \vec x, \psi, \I, \Gamma )$\hspace{1em} where $c \in \{
  \vspace{1ex}
  \cfgmval, \cfgkeep, \cfgeqs, \cfgeqb \}$
  \begin{enumerate}[leftmargin=1em]
    \item[\ ] Let $M = \{ \ell \mid \I \models \ell, \ell \in \lits{\psi} \}$, 
    $N = \{ \project{c}( \I, \ell ) \mid \ell \in M \}$.
    \item[\ ] For $i=1,\ldots,n$ where $\vec x = ( x_1, \ldots, x_n )$:
    \begin{enumerate}
      \item[\ ] 
      Let $N_i = \bigcup_{\ell[ x_1, \ldots, x_{i-1} ] \in N} \linearize{x_i}{\I}{\ell[ t_1, \ldots, t_{i-1} ] }$.
      \item[\ ] 
      Let
      $
      t_i =
      \begin{cases}
        \solvelit{ x_i, \choosef(N_i) } & 
        \text{if } N_i \text{ is non-empty}\\
        x_i^\I & \text{otherwise}
      \end{cases}
      $
      \item[\ ]  $t_j \coloneqq t_j \{ x_i \mapsto t_i \}$ for each $j<i$.
    \end{enumerate}
    \item[\ ] Return $( t_1, \ldots, t_n )$.
  \end{enumerate}
  \vspace{2ex}
  {%
    \renewcommand{\arraystretch}{1.2}%
    \begin{tabular}{ll@{\hskip 1.5em}ll}
      $\project{\cfgmval}( \I, s \rel t )$ &:
        return $\top$ &
      $\project{\cfgeqs}( \I, s \rel t )$ &:
        return $s \teq t + ( s - t )^\I$ \\
      $\project{\cfgkeep}( \I, s \rel t )$ &:
        return $s \rel t$ &
      $\project{\cfgeqb}( \I, s \rel t )$ &:
        return
        $
        \begin{cases}
          s \teq t &
            \text{if } s^\I = t^\I \\
          s \teq t + 1 &
            \text{if } s^\I > t^\I \\
          s \teq t - 1 &
            \text{if } s^\I < t^\I \\
        \end{cases}
        $
    \end{tabular}
  }
\end{framed}

  \vspace{-2ex}
  \caption{Selection functions $\selfunbv{c}$ for quantifier-free bit-vector formulas.
  The procedure is parameterized by a configuration $c$, 
  one of either $\cfgmval$ (model value), $\cfgkeep$ (keep), $\cfgeqs$ (slack), 
  or $\cfgeqb$ (boundary). 
  \label{fig:sel-bv}}
\end{figure}

\subsection{Selection functions for bit-vectors}

In Figure~\ref{fig:sel-bv}, we define a (class of) selection
functions $\selfunbv{c}$ for quantifier-free bit-vector formulas,
which is parameterized by a \emph{configuration} $c$, 
a value of the enumeration type \{\cfgmval, \cfgkeep, \cfgeqs, $\cfgeqb\}$.
The selection function collects in the set $M$ all the literals occurring 
in $\Gamma'$ that are satisfied by $\I$. 
Then, it collects in the set $N$ a \emph{projected form} of each literal in $M$.
This form is computed by the function $\project{c}$ parameterized by configuration $c$. 
That function transforms its input literal into a form suitable 
for function $\solvelitf$ from Figure~\ref{fig:solve}.
We discuss the intuition for projection operations in more detail below.

\begin{rep}
\begin{example}
Consider the $\Sigma_{BV}$-literal $\bvuge{a}{b}$
and the interpretation $\I$ where $a^\I = 5$ and $b^\I = 3$.
We have that $\project{c}$ returns
$\top$ for $c = \cfgmval$,
$\bvuge{a}{b}$ for $c = \cfgkeep$,
$a \teq b + ( 5 - 3 )$ for $c = \cfgeqs$, and
$a \teq b + 1$ for $c = \cfgeqb$.
\exqed
\end{example}

\noindent 
\end{rep}
After constructing set $N$,
the selection function computes a term $t_i$ for each variable $x_i$ in tuple \vec{x},
which we call the \emph{solved form} of $x_i$.
To do that, it first constructs a set of literals $N_i$ all linear in $x_i$.
It considers literals 
$\ell$ from $N$ and replaces
all previously solved variables $x_1, \ldots, x_{i-1}$ by their respective
solved forms to obtain the literal $\ell' = \ell[ t_1, \ldots, t_{i-1} ]$.
It then calls function $\linearizef$ on literal $\ell'$
which returns a \emph{set} of literals, each obtained by 
replacing all but one occurrence of $x_i$ in $\ell$ with
the value of $x_i$ in $\I$.\footnote{
This is a simple heuristic to generate literals that can be solved 
for $x_i$. More elaborate heuristics could be used in practice.
}

\begin{example}
Consider an interpretation $\I$ where $x^\I = 1$,
and $\Sigma_{BV}$-terms $a$ and $b$ with $x \not\in \fv{a} \cup \fv{b}$.
We have that $\linearize{x}{\I}{x \bvmulf ( x + a ) \teq b}$
returns the set $\{ 1 \bvmulf ( x + a ) \teq b, x \bvmulf ( 1 + a ) \teq b \}$;
$\linearize{x}{\I}{ \bvuge{x}{a}}$ returns the singleton
set $\{ \bvuge{x}{a} \}$;
$\linearize{x}{\I}{a \tneq b}$ returns the empty set.
\exqed
\end{example}

\noindent
If the set $N_i$ is non-empty, the selection function
heuristically chooses a literal from $N_i$ 
(indicated in Figure~\ref{fig:sel-bv} with $\choosef(N_i)$).
It then computes a solved form $t_i$ for $x_i$ by solving the chosen literal for $x_i$ 
with the function $\solvelitf$ described in the previous section.
If $N_i$ is empty, we let $t_i$ is simply the value of $x_i$ in the given model $\I$.
After that, $x_i$ is eliminated from all the previous terms $t_1, \ldots, t_{i-1}$ 
by replacing it with $t_i$.
After processing all $n$ variables of $\vec x$,
the tuple $( t_1, \ldots, t_n )$ is returned.

The configurations of selection function \selfunbv{c} determine
how literals in $M$ are modified by the \project{c} function prior to computing
solved forms, based on the current model $\I$.
With the \emph{model value} configuration $\cfgmval$,
the selection function effective ignores the structure of all literals in $M$ and
(because the set $N_i$ is empty)
ends up choosing the value $x_i^\I$ as the solved form variable $x_i$, for each $i$.
On the other end of the spectrum, the configuration~$\cfgkeep$ 
\emph{keeps} all literals in $M$ unchanged.
The remaining two configurations have an effect 
on how disequalities and inequalities are handled by \project{c}.
With configuration $\cfgeqs$ 
\project{c} normalizes any kind of literal (equality, inequality or disequality)
$s \rel t$ to an equality by adding the \emph{slack} value $(s - t)^\I$ to $t$.
With configuration $\cfgeqb$ it maps equalities to themselves 
and inequalities and disequalities 
to an equality corresponding to a \emph{boundary point} of the relation 
between $s$ and $t$ based on the current model.
Specifically, it adds one to $t$ if $s$ is greater than $t$ in $\I$, it subtracts one 
if $s$ is smaller than $t$, and returns $s \teq t$ if their value is the same.
These two configurations are inspired by  
quantifier elimination techniques for 
linear arithmetic~\cite{cooper1972,Loos93applyinglinear}.
In the following, we provide an end-to-end example of
our technique for quantifier instantiation that makes use
of selection function \selfunbv{c}\kern -.5em.

\begin{example}
Consider
formula 
$\varphi = \binde{\vec y}{\binda{x_1}{(x_1 \bvmulf a \bvulef b)}}$
where $a$ and $b$ are terms with no free occurrences of $x_1$.
To determine the satisfiability of $\varphi$, we invoke
$\cegqi{\selfunbv{c}}$ on $\varphi$ for some configuration~$c$.
Say that in the first iteration of the loop, we find that
$\Gamma' = \Gamma \cup \{ x_1 \bvmulf a\bvugtf b \}$ is satisfied
by some model $\I$ of \tbv such that
$x_1^\I = 1$, $a^\I = 1$, and $b^\I = 0$.
We invoke $\selfunbv{c}((x_1), \I, \Gamma')$ and first
compute $M = \{ x_1 \bvmulf a \bvugtf b \}$,
the set of literals of $\Gamma'$ that are satisfied by $\I$.
The table below summarizes the values of the internal variables of 
$\selfunbv{c}$ for the various configurations:
\[
\begin{array}{c@{\hspace{1.5em}}c@{\hspace{1.5em}}c}
  \hline
  \text{config} & N_1 & t_1 \\
  \hline
  \cfgmval & 
  \emptyset & 1 
  \\
  \cfgkeep & 
  \{ x_1 \bvmulf a\bvugtf b \} & 
  \choice{z}{(\bvult{a}{\bvor{\bvneg{b}}{b}}) \limpl z \bvmulf a \bvugtf b}
  \\
  \cfgeqs, \cfgeqb & 
  \{ x_1 \bvmulf a\teq b+1 \} & 
  \choice{z}{(\equal{\bvand{(\bvor{\bvneg{a}}{a})}{b+1}}{b+1}) \limpl z \bvmulf a \teq b+1}
  \\
  \hline
\end{array}
\]
In each case, $\selfunbv{c}$ returns the tuple $( t_1 )$, and
we add the instance $t_1 \bvmulf a \bvulef b$ to $\Gamma$.
Consider configuration $\cfgkeep$ where $t_1$ is the choice expression 
$\choice{z}{((\bvult{a}{\bvor{\bvneg{b}}{b}}) \limpl z \bvmulf a \bvugtf b)}$.
Since $t_1$ is $\choicef$-valid, due to the semantics of $\choicef$, 
this instance is equivalent to:
\begin{align}
\label{eqn:cegqi-instance}
((\bvult{a}{\bvor{\bvneg{b}}{b}}) \limpl k \bvmulf a \bvugtf b) \land k \bvmulf a \bvulef b
\end{align}
for fresh variable $k$.
This formula is \tbv-satisfiable if and only if 
$\neg (\bvult{a}{\bvor{\bvneg{b}}{b}})$
is \tbv-satisfiable.
In the second iteration of the loop in $\cegqi{\selfunbv{c}}$,
set $\Gamma$ contains formula (\ref{eqn:cegqi-instance}) above.
We have two possible outcomes:

$i$) $\neg (\bvult{a}{\bvor{\bvneg{b}}{b}})$ is \tbv-unsatisfiable.
Then (\ref{eqn:cegqi-instance}) and hence $\Gamma$ are \tbv-unsatisfiable, 
and the procedure terminates with ``unsat''.

$ii$) $\neg (\bvult{a}{\bvor{\bvneg{b}}{b}})$ is satisfied by some model $\J$ of \tbv.
Then $\exists z. \bvule{\bvmul{z}{a}}{b}$ is false in $\J$
since the  invertibility condition of $\bvule{\bvmul{z}{a}}{b}$ is false in $\J$.
Hence, $\Gamma' = \Gamma \cup \{ x_1 \bvmulf a\bvugtf b \}$
is unsatisfiable, and the algorithm terminates with ``sat''.

In fact,
we argue later that quantified bit-vector formulas like $\varphi$ above, which
contain only one occurrence of a universal variable,
require at most one instantiation before $\cegqi{\selfunbv{\cfgkeep}}$ terminates.
The same guarantee does not hold with the other configurations.
In particular, 
configuration $\cfgmval$ generates 
the instantiation where $t_1$ is $1$, which simplifies to $a \bvulef b$.
This may not be sufficient to show that $\Gamma$ or $\Gamma'$ is unsatisfiable
in the second iteration of the loop
and
the algorithm may resort to \emph{enumerating} a repeating pattern of 
instantiations, such as $x_1 \mapsto 1, 2, 3, \ldots$ and so on.
This obviously does not scale for problems with large bit-widths.
\exqed
\end{example}

\begin{rep}
\begin{example}
As the last example demonstrates,
$\cegqi{\selfunbv{\cfgkeep}}$ may terminate with at most one instance
for input formulas whose body has just one literal and a single occurrence 
of each universal variable.
However, consider extending the quantified formula
from the previous example to a disjunction of two literals:
$\binde{\vec{y} \forall x_1}{(x_1 \bvmulf a \bvulef b \lor \ell[x_1])}$.
Assume that our selection function chooses the same $t_1$ as in the previous example.
The corresponding instance is equivalent to:
\begin{align}
\label{eqn:cegqi-instance-mul}
((\bvult{a}{\bvor{\bvneg{b}}{b}}) \limpl k \bvmulf a \bvugtf b) \land 
( k \bvmulf a \bvulef b \lor \ell[k] )
\end{align}
In contrast to the previous example,
the second iteration of the loop from Figure~\ref{fig:proc-qi} 
is not guaranteed to terminate for this example.
The above formula may be satisfied by a model $\J$
where $k \bvmulf a \bvugtf b$ and $\ell[k]$ hold.
Notice that $\J$ may also satisfy $\bvult{a}{\bvor{\bvneg{b}}{b}}$,
meaning it may still be the case that 
$x_1 \bvmulf a \bvulef b$ together with the above instance is satisfied by $\J$.
In such a case, we may invoke $\cegqi{\selfunbv{\cfgkeep}}$ again,
which may produce the same solved form for $x_1$ if
it constructs a solved form for $x_1$ 
again based on the literal $x_1 \bvmulf a \bvulef b$.
Hence, by the terminology from Definition~\ref{def:selfun-prop},
the selection function $\selfunbv{\cfgkeep}$ is not monotonic for
quantified formulas with more than one occurrence of a universal variable.
\exqed
\end{example}
\end{rep}\begin{conf}
\noindent
More generally, we note that
$\cegqi{\selfunbv{\cfgkeep}}$ terminates with at most one instance
for input formulas whose body has just one literal and a single occurrence 
of each universal variable.
The same guarantee does not hold for instance for quantified formulas
whose body has multiple disjuncts.
For some intuition, consider extending the second conjunct of 
(\ref{eqn:cegqi-instance}) with an additional disjunct, i.e.
$( k \bvmulf a \bvulef b \lor \ell[k] )$.
A model can be found for this formula 
in which the invertibility condition $(\bvult{a}{\bvor{\bvneg{b}}{b}})$
is still satisfied, and hence we are not guaranteed to terminate on the 
second iteration of the loop.
\end{conf}
Similarly,
if the literals of the input formula have multiple 
occurrences of $x_1$, then multiple
instances may be returned by the selection function
since the literals returned
by $\linearizef$ in Figure~\ref{fig:sel-bv} 
depend on the model value of $x_1$, and hence more than one possible
instance may be considered in loop in Figure~\ref{fig:proc-qi}.

The following theorem summarizes the properties
of our selection functions.
In the following, we say a quantified formula
is \emph{unit linear invertible} if it is of the form $\forall x. \ell[x]$
where $\ell$ is linear in $x$
and has an invertibility condition for $x$.
We say a selection function is \emph{$n$-finite} 
for a quantified formula $\psi$
if the number of possible instantiations 
it returns is at most $n$ for some positive integer $n$.

\begin{theorem}
\label{thm:sel-bv-prop}
Let $\psi[\vec x]$ be a quantifier-free formula in the signature of \tbv.
\begin{enumerate}
\item 
$\selfunbv{c}$ is a finite selection function for $\vec x$ and $\psi$
for all $c \in \{\cfgmval, \cfgkeep, \cfgeqs, \cfgeqb\}$.
\item 
$\selfunbv{\cfgmval}$ is monotonic.
\item 
$\selfunbv{\cfgkeep}$ is $1$-finite if $\psi$ is unit linear invertible.
\item $\selfunbv{\cfgkeep}$ is monotonic if $\psi$ is unit linear invertible.
\end{enumerate}
\end{theorem}
\begin{rep}
\begin{proof}
  Let \vec{x = ( x_1, \ldots, x_n )} be a tuple of variables
and let $\varphi$ be a quantifier-free $T_{BV}$-formula.
We show each part for the case where $n=1$; the arguments 
below can be lifted to $n>1$ in a straightforward way.
Let $\Gamma$ be a set of formulas such that $x_1 \not \in \fv{\Gamma}$,
let $\I$ be a model of $T_{BV}$ such that 
$\I \models \Gamma \cup \{ \neg \varphi \}$,
and let $t_1$ be $\selfunbv{c}( (x_1), \I, \Gamma )$.

Part 1)
To show that $\selfunbv{c}$ is a selection function,
we must show that $t_1$ is $\choicef$-valid and 
$\fv{t_1} \subseteq \fv{\varphi} \setminus \{ x_1 \}$.
Notice that for all configurations,
the value $( t_1 )$ returned by $\selfunbv{c}$ is either 
of the form $x_1^\I$ or $\solvelitf( x_1, \ell )$ 
for some $\ell \in \lits{\Gamma}$.
In the former case, we have that $t_1$ is clearly $\choicef$-valid 
and $\fv{t_1} = \emptyset$.
In the latter case, as a consequence of Theorem~\ref{thm:solve},
we have that $t_1$ is $\choicef$-valid and $\fv{t_1} \subseteq \fv{\ell}$.
We have that $\ell \in \lits{\varphi}$
and thus $\fv{\ell} \subseteq \fv{\varphi}$.
Thus, in either case, 
we have $\fv{t_1} \subseteq \fv{\varphi} \setminus \{ x_1 \}$.
Hence, $\selfunbv{c}$ is a selection function for 
$c=\cfgmval$, $\cfgkeep$, $\cfgeqs$, $\cfgeqb$.
To show these selection functions are finite for $\varphi$,
note that the number of terms of form $x_1^\I$ is finite.
Also note that the number of literals in $\lits{\varphi}$ is finite.
For each $c$ and $\ell \in \lits{\varphi}$, 
set of literals of the form $\project{c}( \I, \ell )$, 
call this set $N$, is finite.
For each literal $\ell' \in N$, the set of literals
returned by $\linearizef( x_i, \I, \ell' )$, call this set $N'$,
is also finite.
Since it is either the case that 
$t_1 = \solvelitf( x_1, \ell'' )$ for some $\ell'' \in N'$
or $t_1 = x_1^\I$, the number of possible return values
of $\selfunbv{c}$ is finite for $( x_1 )$ and $\varphi$
for $c=\cfgmval$, $\cfgkeep$, $\cfgeqs$, $\cfgeqb$.

Part 2)
Since $x_1 \not\in \fv{\ell}$ for any literal returned by
$\project{\cfgmval}( \I, \ell )$, it must be the case that $t_1 = x_1^\I$.
Assume that $\project{\cfgmval}$ was not monotonic
for $( x_1 )$ and $\varphi = \neg \psi[ x_1 ]$.
Since $\project{\cfgmval}$ is a selection function by Part 1
and $\I, \Gamma$ is a legal input to $\project{\cfgmval}$,
then without loss of generality we may assume that
$\psi[ t_1 ] \in \Gamma$.
However, 
$\neg \psi[ x_1 ] \in \Gamma$, $\I \models \Gamma$ and since
$t_1^\I = x_1^\I$, 
it must instead be the case that $\varphi[ t_1 ] \not\in \Gamma$.
Hence, $\project{\cfgmval}$ is monotonic for $\varphi$.

Part 3)
Assume $\varphi$ is the unit linear invertible
formula $\ell$.
By definition of $\project{\cfgkeep}$, $\linearizef$,
and since by definition $\ell$ is linear with respect to $x_1$
and $\I \models \neg \ell$,
we have that $t_1$ must
be the term returned by $\solvelitf( x_1, \neg \ell )$.
Hence, $\selfunbv{\cfgkeep}$ has only one possible return value
and hence is $1$-finite.

Part 4)
Assume $\varphi$ is a unit linear invertible
formula $\ell[ x_1 ]$.
The return value of $\selfunbv{\cfgkeep}$ is the tuple $( t_1 )$,
where by the reasoning in Part 3, we have that $t_1$ is the term returned by
$\solvelitf( x_1, \neg \ell[ x_1 ] )$.
Thus by Theorem~\ref{thm:solve},
we have that 
$\neg \ell[t_1] \lequiv \binde{z}{\neg \ell[z]}$ holds in all models of \tbv.
Now, assume that $\project{\cfgkeep}$ was not monotonic 
for $( x_1 )$ and $\ell[ x_1 ]$.
Since $\project{\cfgkeep}$ is a selection function by Part 1
and $\I, \Gamma$ is a legal input to $\project{\cfgkeep}$,
then without loss of generality we may assume that $\ell[ t_1 ] \in \Gamma$.
Since $\I \models \Gamma$,
and since 
$\neg \ell[t_1] \Leftarrow \binde{z}{\neg \ell[z]}$ 
holds in all models of \tbv,
we have that $\I$ must satisfy $\neg \binde{z}{\neg \ell[z]}$,
which is $\binda{z}{\ell[z]}$.
However, we also have that by assumption $\neg \ell[ x_1 ] \in \Gamma$.
Hence, 
it must instead be the case that $\varphi[ t_1 ] \not\in \Gamma$
and thus $\project{\cfgkeep}$ is monotonic for $\varphi$.
\qed

\end{proof}
\end{rep}

\noindent
This theorem implies
that counterexample-guided instantiation using 
configuration~$\selfunbv{\cfgmval}$ is a decision
procedure for quantified bit-vectors.
However, in practice the worst-case
number of instances considered by this configuration 
for a variable $\bv{x}{n}$ is proportional to the number of its possible values
($2^n$), which is practically infeasible for sufficiently large $n$.
More interestingly, counterexample-guided instantiation using $\selfunbv{\cfgkeep}$
is a decision procedure for quantified formulas that are unit linear invertible,
and moreover has the guarantee that at most one instantiation is returned
by this selection function. Hence, 
formulas in this fragment can be effectively reduced to
quantifier-free bit-vector constraints in at most two iterations of the loop
of procedure \cegqi{\selfun} in Figure~\ref{fig:proc-qi}.

\subsection{Implementation}
\label{sec:implementation}

We implemented 
the new instantiation techniques described in this section
as an extension of \cvc,
which is
a \dpllt-based SMT solver~\cite{NieOT-JACM-06}
that supports quantifier-free bit-vector constraints, 
(arbitrarily nested) quantified formulas, and 
support for choice expressions.
For the latter, all choice expressions $\choice{x}{\varphi[x]}$ are 
eliminated from assertions by replacing them with a fresh variable $k$
of the same type and adding $\varphi[k]$ as a new assertion,
which notice is sound since all choice expressions we consider are $\varepsilon$-valid.
In the remainder of the paper, we will refer to 
our extension of the solver as \solver.
In the following, we discuss important implementation details
of the extension.
\medskip

\noindent\emph{Handling Duplicate Instantiations}
The selection functions
$\selfunbv{\cfgeqs}$ and $\selfunbv{\cfgeqb}$ are not guaranteed to be monotonic,
neither is $\selfunbv{\cfgkeep}$
for quantified formulas that contain more than one occurrence
of universal variables.
Hence, when applying these strategies to arbitrary quantified formulas,
we use a two-tiered strategy
that invokes $\selfunbv{\cfgmval}$ as a second resort if the instance for 
the terms returned by a selection function already exists in $\Gamma$.
\medskip

\noindent\emph{Linearizing Rewrites}
Our selection function in Figure~\ref{fig:sel-bv}
uses the function $\linearizef$
to compute literals that are linear in the variable $x_i$ to solve for.
The way we presently implement $\linearizef$ makes those literals
dependent on the value of $x_i$ in the current model $\I$,
with the risk of overfitting to that model.
To address this limitation, we use a set of equivalence-preserving 
rewrite rules whose goal is to reduce the number of occurrences of $x_i$
to one when possible,
by applying basic algebraic manipulations.
As a trivial example, a literal like
$x_i \bvaddf x_i \teq a$ is rewritten first to $2 \bvmulf x_i \teq a$
which is linear in $x_i$ if $a$ does not contain $x_i$.
In that case, this literal, and so the original one, 
has an invertibility condition as discussed in Section~\ref{sec:bvinversion}.

\medskip

\noindent\emph{Variable Elimination}
We use procedure \solvelitf from Section~\ref{sec:bvinversion}
not only for selecting quantifier instantiations,
but also for eliminating variables from quantified formulas.
In particular, for a quantified formula of the form 
$\binda{x \vec{y}}{\ell \limpl \varphi[x, \vec{y} ]}$,
if $\ell$ is linear in $x$ and $\solvelitf( x, \ell )$
returns a term $s$ containing no $\choicef$-expressions,
we can replace this formula by
$\binda{\vec{y}}{\varphi[s,\vec{y} ]}$.
When $\ell$ is an equality, this is sometimes called
destructive equality resolution (DER)
and is an important implementation-level optimization in state-of-the-art
bit-vector solvers~\cite{WintersteigerHM13}.
\begin{conf}
As shown in Figure~\ref{fig:solve},
we use the $\getinversef$ function
to increase the likelihood that $\solvelitf$ returns a term 
that contains no $\choicef$-expressions.
\end{conf}
\begin{rep}
In our approach,
to increase the likelihood that $\solvelitf$ returns a term 
that contains no $\choicef$-expressions, we include several optimizations that
determine when it can be determined that $\ell$ has a unique solution
for $x$. A common example is an equality that involves
multiplication by an odd constant, i.e. $x \bvmulf c \teq t$ where 
$c$ is an odd constant. The only solution for $x$ in this case is
$c^{-1} \bvmulf t$ where $c^{-1}$ denotes the (unique) multiplicative
inverse of $c$ modulo the bit-width of the type of $c$, which can be computed
by Euclid's algorithm.
\end{rep}
\medskip

\noindent\emph{Handling Extract}
Consider formula 
$\binda{\bv{x}{32}}{(\bvextract{x}{31}{16} \tneq \bv{a}{16} \lor 
\bvextract{x}{15}{0} \tneq \bv{b}{16})}$.
Since all invertibility conditions for the extract operator are \true,
rather than producing choice expressions
we have found it more effective to eliminate extracts via rewriting.
As a consequence,
we independently solve 
constraints for \emph{regions}
of quantified variables when they appear underneath
applications of extract operations.
In this example, we let the solved form of $x$ be 
$\bv{y}{16} \bvconcatf \bv{z}{16}$ where $y$ and $z$ are fresh variables,
and subsequently solve for these variables in
$\equal{y}{a}$ and $\equal{z}{b}$.
Hence, we may instantiate
$x$ with $a \bvconcatf b$, a term that we would not have found
by considering the two literals independently in the negated body
of the formula above.

\begin{rep}
\medskip

\noindent\emph{Handling Propositional Structure and Nested Quantifiers}
Notice that Figure~\ref{fig:proc-qi}
describes counterexample-guided quantifier instantiation 
for an input formula with one level of quantifier alternation.
In practice, our technique can be used for problems
containing more than one level of quantifier alternation and that are 
not in prenex normal form.
A thoroughout description of this technique is beyond the scope of the paper;
we provide some high level details here.
Recall that in the \dpllt setting,
the SMT solver incrementally builds a truth assignment
in the form of a set of literals, with the goal of finding a set
that propositionally satisfies all clauses in $\Gamma$ 
and is consistent with respect to the background theory.
In this setting, we consider all quantified formulas 
$\forall \vec{x}. \varphi[ \vec{x} ]$ in the current set $M$.
For each of these formulas, we may add clauses of three forms:
\emph{instantiation lemmas} of the form 
$( \neg A \lor \varphi[ \vec{t} ] )$,
\emph{Skolemization lemmas} of the form 
$( \neg B \lor \neg \varphi[ \vec{k} ] )$ where 
\vec{k} are fresh constants of the same type as \vec{x}, and 
the connecting clauses
$(\forall \vec{x}. \varphi[ \vec{x} ] \limpl A)$ and
$(\neg \forall \vec{x}. \varphi[ \vec{x} ] \limpl B)$
Here, $A$ and $B$ are fresh Boolean constants which we
call the \emph{positive} and \emph{negative} guard of 
$\forall \vec{x}. \varphi[ \vec{x} ]$ respectively.
The second and third clauses are added once at the time when 
the quantified formula first occurs in an assignment $M$.
We detect when the negation
of a quantified formula along with the current set of clauses $\Gamma$ 
is unsatisfiable by checking which negative guards $B$ must be assigned to false.
In practice, this is determined by a decision heuristic which insists
that negative guards must be decided with positive polarity first.
If a quantified formula and its corresponding negative guard are both
asserted true, then we add an instantiation lemma to $\Gamma$ based on the 
selection function from Figure~\ref{fig:sel-bv}.
We terminate as usual when the set $\Gamma$ is unsatisfiable,
or we find a consistent satisfying assignment where no quantified formula
and its negative guard are both asserted.
This scheme allows the SMT solver to handle multiple quantified formulas
simultaneously, as well as handling quantified formulas with arbitrary nesting.
Above, notice that $\varphi[ \vec{k} ]$ may contain quantifiers,
which are recursively handled by introducing instantiation and Skolemization
lemmas for quantified formulas that appear in subsequent satisfying assignments.
\medskip

\noindent\emph{Negating the Input Formula}
Our version of counterexample-guided quantifier instantiation is most effective 
for input formulas that are closed and universal.
Thus, when an input formula is of the form \binde{\vec{x}}{\varphi} 
where \vec{x} is non-empty and $\varphi$ may contain quantifiers,
we consider its negation \binda{\vec{x}}{\neg \varphi} instead. 
The latter formula may be significantly easier to solve since 
our quantifier instantiation techniques may find an instantiation
for~\vec{x} that quickly leads to a proof of unsatisfiability,
whereas instantiating \vec{x} is not possible for the former.
Since the theory of bit-vectors is a complete theory, 
it follows that
the original formula is satisfiable if and only if this formula
is unsatisfiable.
\medskip

\noindent\emph{Rewrite Rules for Quantifier-Free Constraints}
Finally, we have found that in a \dpllt-based SMT solver, the quantifier-free
bit-vector solver is often the bottleneck 
when solving quantified bit-vector constraints.
For this reason, we use aggressive rewriting techniques for 
quantifier-free bit-vector constraints with the goal
of replacing constraints with expensive propositional encodings with those
with simpler encodings.
\end{rep}

\section{Evaluation}

  We implemented our techniques
  in the solver \solver
  and considered four configurations
  \solvercfg{c},
  where \textbf{c} is one of \{\cfgmval, \cfgkeep, \cfgeqs, \cfgeqb\},
  corresponding to the four selection function configurations described
  in Section~\ref{sec:cegqibv}.
  Out of these four configurations,
  \solvercfg{\cfgmval}
  is the only one
  that does not employ our new techniques
  but uses only model values for instantiation.
  It can thus be considered our base configuration.
  All configurations enable
  the optimizations described in Section~\ref{sec:implementation}
  when applicable.
  We compared them
  against
  all entrants
  of the quantified bit-vector division
  of the 2017 SMT competition SMT-COMP:
  \boolector~\cite{boolector2014},
  \cvc~\cite{CVC4},
  \qiiib~\cite{q3b2016}
  and \ziii~\cite{Z3}.
  With the exception of \qiiib, all solvers are related to our approach
  since they are instantiation-based.
  However, none of these solvers
  utilizes invertibility conditions when constructing instantiations.
  We ran all experiments on the StarExec logic solving service~\cite{Stump2014}
  with a 300 second CPU and wall clock time limit and 100 GB memory limit.

  \begin{table}[t]
    \centering
    \scalebox{1.0}{{%
  \begin{tabular}{@{}|@{\hskip 0.5em}c@{\hskip 0.5em}|@{\hskip 0.5em}cccccccc@{\hskip 0.5em}|@{}}
    \hline
    \rule{0pt}{2.4ex}
    \bf{unsat} & \bf{\boolector} & \bf{\cvc} & \bf{\qiiib} & \bf{\ziii} & 
    \bf{\solvercfg{\cfgmval}} & \bf{\solvercfg{\cfgkeep}} & 
    \bf{\solvercfg{\cfgeqs}} & \bf{\solvercfg{\cfgeqb}}
    \\[.1ex]
    \hline
    \rule{0pt}{2.4ex}
    { h-uauto } & 14 & 12 & 93 & 24 & 10 & 103 & 105 & \bf{ 106 } \\
    { keymaera } & 3917 & 3790 & 3781 & \bf{ 3923 } & 3803 & 3798 & 3888 & 3918 \\
    { psyco } & \bf{ 62 } & \bf{ 62 } & 49 & \bf{ 62 } & \bf{ 62 } & 39 & \bf{ 62 } & 61 \\
    { scholl } & 57 & 36 & 13 & \bf{ 67 } & 36 & 27 & 36 & 35 \\
    { tptp } & 55 & 52 & \bf{ 56 } & \bf{ 56 } & \bf{ 56 } & \bf{ 56 } & \bf{ 56 } & \bf{ 56 } \\
    { uauto } & \bf{ 137 } & 72 & 131 & \bf{ 137 } & 72 & 72 & 135 & \bf{ 137 } \\
    { ws-fixpoint } & 74 & 71 & \bf{ 75 } & 74 & \bf{ 75 } & 74 & \bf{ 75 } & \bf{ 75 } \\
    { ws-ranking } & 16 & 8 & 18 & \bf{ 19 } & 15 & 11 & 12 & 11 
    \\[.1ex]
    \hline
    \rule{0pt}{2.4ex}
    \bf{Total unsat} & 4332 & 4103 & 4216 & 4362 & 4129 & 4180 & 4369 & \bf{ 4399 } 
    \\[.1ex]
    \hline
    \hline
    \rule{0pt}{2.4ex}
    \bf{sat} & \bf{\boolector} & \bf{\cvc} & \bf{\qiiib} & \bf{\ziii} & 
    \bf{\solvercfg{\cfgmval}} & \bf{\solvercfg{\cfgkeep}} & 
    \bf{\solvercfg{\cfgeqs}} & \bf{\solvercfg{\cfgeqb}}
    \\[.1ex]
    \hline
    { h-uauto } & 15 & 10 & \bf{ 17 } & 13 & 16 & \bf{ 17 } & 16 & \bf{ 17 } \\
    { keymaera } & \bf{ 108 } & 21 & 24 & \bf{ 108 } & 20 & 13 & 36 & 75 \\
    { psyco } & 131 & \bf{ 132 } & 50 & 131 & \bf{ 132 } & 60 & \bf{ 132 } & 129 \\
    { scholl } & \bf{ 232 } & 160 & 201 & 204 & 203 & 188 & 208 & 211 \\
    { tptp } & \bf{ 17 } & \bf{ 17 } & \bf{ 17 } & \bf{ 17 } & \bf{ 17 } & \bf{ 17 } & \bf{ 17 } & \bf{ 17 } \\
    { uauto } & 14 & 14 & 15 & \bf{ 16 } & 14 & 14 & 14 & 14 \\
    { ws-fixpoint } & 45 & 49 & \bf{ 54 } & 36 & 45 & 51 & 49 & 50 \\
    { ws-ranking } & 19 & 15 & \bf{ 37 } & 33 & 33 & 31 & 31 & 32 
    \\[.1ex]
    \hline
    \rule{0pt}{2.4ex}
    \bf{Total sat} & \bf{ 581 } & 418 & 415 & 558 & 480 & 391 & 503 & 545 
    \\[.1ex]
    \hline
    \hline
    \rule{0pt}{2.4ex}
    \bf{Total (5151)} & 4913 & 4521 & 4631 & 4920 & 4609 & 4571 & 4872 & \bf{ 4944 } 
    \\[.1ex]
    \hline
  \end{tabular}
}

}
    \vspace{1ex}
    \caption{Results on satisfiable and unsatisfiable benchmarks
    with a 300 second timeout.
    \label{tab:results-smtlib}}
  \end{table}

  We evaluated our approach on all 5,151 benchmarks
  from the quantified bit-vector logic~(BV) of SMT-LIB~\cite{SMTLib2010}.
  The results are summarized in Table~\ref{tab:results-smtlib}.
  Configuration \solvercfg{\cfgeqb}
  solves the highest number of unsatisfiable benchmarks ($4,399$),
  which is $30$ more than the next best configuration
  \solvercfg{\cfgeqs}
  and $37$ more than the next best external solver, \ziii.
  Compared to the instantiation-based solvers
  \boolector, \cvc and \ziii,
  the performance of \solvercfg{\cfgeqb} is particularly strong
  on the {h-uauto} family, which are verification conditions
  from the Ultimate Automizer tool~\cite{HeizmannCDGNMSS17}.
  For satisfiable benchmarks, Boolector solves the most (581),
  which is $36$ more than our best configuration
  \solvercfg{\cfgeqb}.

\begin{wrapfigure}{r}{0.5\textwidth}
    \vspace{-6ex}
    \centering
    \scalebox{0.33}{%
      \includegraphics[trim={0 0 0 0}, clip]{%
        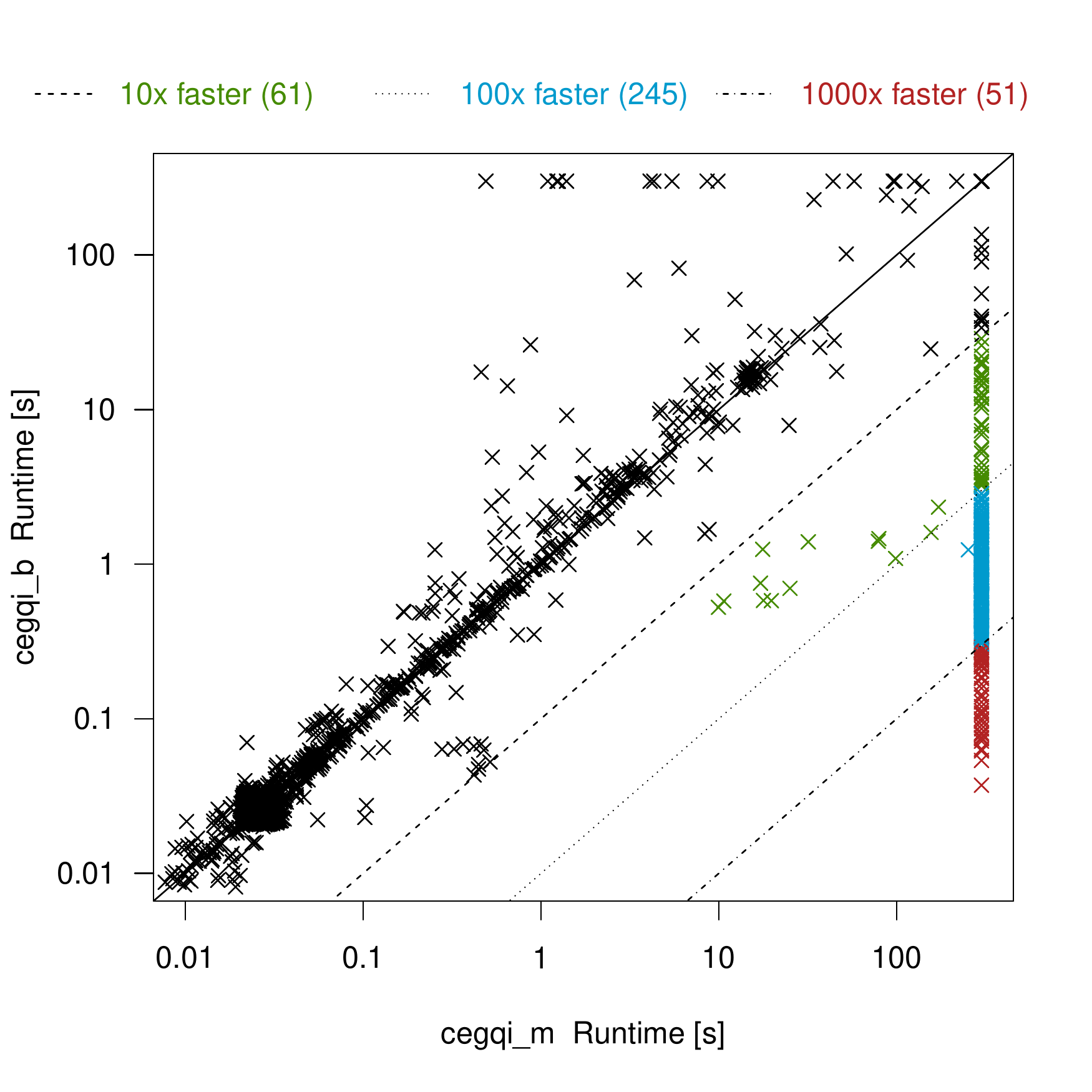}}
    \vspace{-3ex}
    \caption{Configuration \solvercfg{\cfgmval} vs.~\solvercfg{\cfgeqb}.}
    \label{fig:scatter}
    \vspace{-4ex}
  \end{wrapfigure}

  Overall,
  our best configuration \solvercfg{\cfgeqb}
  solved $335$ more benchmarks than our base configuration
  \solvercfg{\cfgmval}.
  A more detailed runtime comparison between the two is provided by the 
  scatter plot in 
  Figure~\ref{fig:scatter}.
  Moreover,
  \solvercfg{\cfgeqb} solved $24$ more benchmarks than the best external solver,
  \ziii.
  In terms of uniquely solved instances,
  \solvercfg{\cfgeqb} was able to solve 139 benchmarks
  that were not solved by \ziii, whereas \ziii solved 115
  benchmarks that \solvercfg{\cfgeqb} did not.
  Overall, \solvercfg{\cfgeqb} was able to solve 21 of the 79
  benchmarks (26.6\%) not solved by any of the other solvers.
  For 18 of these 21 benchmarks,
  it terminated after 
  considering no more than 4 instantiations.
  These cases indicate that using symbolic terms for instantiation
  solves problems
  for which other techniques,
  such as those that
  enumerate instantiations based on model values,
  do not scale.

  Interestingly,
  configuration \solvercfg{\cfgkeep},
  despite having the strong guarantees given by Theorem~\ref{thm:sel-bv-prop},
  performed relatively poorly on this set (with $4,571$ solved instances overall).
  We attribute this to the fact that most of the quantified formulas
  in this set are not unit linear invertible.
  In total, we found that only 25.6\% of the formulas considered during
  solving were unit linear invertible.
  However, only a handful of benchmarks
  were such that \emph{all} quantified formulas in the problem were
  unit linear invertible. 
  This might explain the superior performance of
  \solvercfg{\cfgeqs} and $\bf{\solvercfg{\cfgeqb}}$
  which use invertibility conditions but in a less monolithic way.

  \noindent
  For some intuition on this, consider the problem 
  \binda{x}{(x > a \lor x < b)}
  where $a$ and $b$ are such that $a>b$ is \tbv-valid.
  Intuitively, to show that this formula is unsatisfiable
  requires the solver to find an $x$
  between $b$ and $a$.
  This is apparent when
  considering the dual problem
  \binde{x}{(x \leq a \land x \geq b)}.
  Configuration \solvercfg{\cfgeqb} is capable of finding
  such an $x$, for instance, by considering the instantiation $x \mapsto a$
  when solving for the boundary point of the first disjunct.
  Configuration \solvercfg{\cfgkeep}, on the other hand,
  would instead consider the 
  instantiation of $x$ for two terms that witness $\choicef$-expressions:
  some $k_1$ that is never smaller than $a$, and 
  some $k_2$ that is never greater that $b$.
  Neither of these terms necessarily resides in between $a$ and $b$
  since
  the solver may subsequently consider models where $k_1 > b$ and $k_2 < a$.
  This points to a potential use for invertibility conditions
  that solve multiple literals simultaneously, 
  something we are currently investigating.

\section{Conclusion}

We have presented a new class of strategies for solving 
quantified bit-vector formulas based on invertibility conditions.
We have derived invertibility conditions for
the majority of operators in a standard theory of 
fixed-width bit-vectors.
An implementation based on this approach solves over 25\% 
of previously unsolved verification benchmarks from SMT LIB, 
and outperforms all other state-of-the-art bit-vector solvers overall.

In future work, 
we plan to develop a framework in which the correctness
of invertibility conditions can be formally established
independently of bit-width.
We are working on 
deriving invertibility conditions that 
are optimal for linear constraints, in the sense of admitting
the simplest propositional encoding.
We also are investigating conditions
that cover additional bit-vector operators,
some cases of non-linear literals, as well as
those that cover multiple constraints. 
While this is a challenging task, we believe efficient
syntax-guided synthesis solvers can continue to help push progress 
in this direction.
Finally, we plan to investigate the use of invertibility conditions
for performing quantifier elimination on bit-vector constraints.
This will require a procedure for deriving
concrete witnesses from choice expressions.

\bibliographystyle{splncs04}
\bibliography{biblio}

\end{document}